\documentclass[12pt]{article}
\pdfoutput=1
\usepackage[nosort]{cite}

\usepackage{draft}

\usepackage{epsfig}
\usepackage{amsmath,amssymb}
\usepackage{cite}
\usepackage{hyperref}
\hypersetup{colorlinks=true}
\hypersetup{linkcolor=black}
\hypersetup{citecolor=black}
\hypersetup{urlcolor=black}
 
\usepackage{tikz}
\usepackage{tikz-cd}

\begin{document}

\begin{titlepage}
	\hfill YITP-SB-2021-29, MIT-CTP/5359

\title{Non-invertible Duality Defects in 3+1 Dimensions}

\author{Yichul Choi${}^{1,2}$, Clay C\'ordova${}^3$, Po-Shen Hsin${}^4$, Ho Tat Lam${}^5$, Shu-Heng Shao${}^1$}

		\address{${}^{1}$C.\ N.\ Yang Institute for Theoretical Physics, Stony Brook University\\
        ${}^{2}$Simons Center for Geometry and Physics, Stony Brook University\\
		${}^{3}$Enrico Fermi Institute $\&$ Kadanoff Center for Theoretical Physics,
University of Chicago
		\\
		${}^{4}$Mani L. Bhaumik Institute for Theoretical Physics,   UCLA		\\
		${}^{5}$Center for Theoretical Physics, Massachusetts Institute of Technology
		}
\abstract

For any quantum system invariant under gauging a higher-form global symmetry, we construct a non-invertible topological defect by gauging in only half of spacetime. This generalizes the Kramers-Wannier duality line in 1+1 dimensions to higher spacetime dimensions. We focus on the case of a one-form symmetry in 3+1 dimensions, and determine the fusion rule. From   a direct analysis of one-form symmetry protected topological phases, we show that the existence of certain kinds of duality defects is intrinsically incompatible with a trivially gapped phase. 
 We give an explicit realization of this duality defect in the free Maxwell theory, both in the continuum and in a modified Villain lattice model. The duality defect is realized by a Chern-Simons coupling between the gauge fields from the two sides.  We further construct the duality defect in non-abelian gauge theories and the $\mathbb{Z}_N$ lattice gauge theory.

\end{titlepage}

\eject

\tableofcontents

\section{Introduction}

Global symmetry is one of the most important tools in analyzing strongly coupled systems. 
In recent years, there has been a revolution in our understanding of symmetries, anomalies, and their generalizations. 
A prominent extension of the familiar notion of ordinary global symmetries is higher-form symmetries \cite{Gaiotto:2014kfa} and their generalized fusion categories described by higher-groups \cite{Kapustin:2013uxa, Cordova:2018cvg, Benini:2018reh}.  Even more broadly the idea of symmetry has been expanded to include non-invertible topological defects \cite{Bhardwaj:2017xup,Chang:2018iay}.  These more general fusion algebras fundamentally link the ideas of symmetry and topological field theory.  In this paper, we discuss an interesting interplay between these notions of higher symmetry and non-invertability in diverse spacetime dimensions.

The modern approach to characterizing global symmetries is via their symmetry  operators or defects. 
For an ordinary global symmetry $G$, the symmetry operator  $U_g$ for any group element $g\in G$ acts on all of space at a given time, and is conserved under time evolution. 
In  relativistic systems, such an operator becomes a codimension-one topological defect $U_g(M^{(d-1)})$ wrapping a closed $(d-1)$-manifold $M^{(d-1)}$ in the $d$-dimensional spacetime $X^{(d)}$.  
The symmetry defects obey the standard group-like fusion rule: $U_{g_1} U_{g_2} =U_{g_1g_2}$.\footnote{Here and in the following, we often leave the manifold wrapped by a defect implicit, with all fusion rules understood as implying that the defect loci of fusing operators coincide.}

However, not every  codimension-one topological defect is associated with an ordinary global symmetry. 
These more general topological defects ${\cal D}(M^{(d-1)})$ do not obey a group-like fusion rule: 
their fusion generally involves more than one defect, or even defects of different dimensions. 
The simplest example of such a defect is the Kramers-Wannier duality defect line in the 1+1d critical Ising conformal field theory (CFT) \cite{Frohlich:2004ef,Frohlich:2006ch,Frohlich:2009gb,Chang:2018iay}, which arises from an anomalous $\mathbb{Z}_2$ symmetry in the Majorana CFT under bosonization \cite{Thorngren:2018bhj,Ji:2019ugf,Lin:2019hks}.

In 1+1d, non-invertible topological lines are ubiquitous.  
In rational CFT \cite{Moore:1988qv,Moore:1989yh,Fuchs:2002cm}, there is  a general construction of topological lines that commute with the extended chiral algebra \cite{Verlinde:1988sn,Petkova:2000ip}.  
In recent years, it has been advocated that these non-invertible topological defects should be viewed as  a generalization of the ordinary global symmetries \cite{Bhardwaj:2017xup,Chang:2018iay}. 
The concept of gauging \cite{Frohlich:2009gb,Carqueville:2012dk,Brunner:2013xna} these lines in 1+1d has been revisited in \cite{Bhardwaj:2017xup,Thorngren:2019iar,Gaiotto:2020iye,Huang:2021zvu}, and the obstruction to gauging can be viewed as a generalized notion of a 't Hooft anomaly. 
Similar to the anomaly constraints from an ordinary symmetry, these non-invertible topological defect lines have dramatic consequences on renormalization group flows. 
In 1+1 dimensional quantum field theory (QFT),  the topological lines have been explored extensively in \cite{Bhardwaj:2017xup,Chang:2018iay,Lin:2019kpn,Thorngren:2019iar,Cordova:2019wpi,Lin:2019hks,Komargodski:2020mxz,Yu:2020twi,Chang:2020imq,Pal:2020wwd,Hegde:2021sdm,Lin:2021udi,Inamura:2021wuo,Grigoletto:2021zyv,Thorngren:2021yso,Ji:2021esj,Huang:2021ytb,Huang:2021zvu} with various dynamical applications.  
In particular, using a modular invariance argument, it was shown in \cite{Chang:2018iay} (see also \cite{Thorngren:2019iar} for generalizations) that the existence of certain non-invertible topological lines is incompatible with a trivially gapped phase. 

Much less is known about non-invertible defects in higher spacetime dimensions. 
In 2+1d topological quantum field theory (TQFT), there are non-invertible surface operators \cite{Kapustin:2010if}. 
The non-abelian anyons in the 2+1d TQFT can also be viewed as  a non-invertible version of the one-form symmetry (see \cite{Kaidi:2021gbs} for an application of this perspective). 
Non-invertible topological defects have been applied to conjectures in  quantum gravity \cite{Rudelius:2020orz,Heidenreich:2020tzg,McNamara:2021cuo}. 
See also \cite{Nguyen:2021yld} for  other discussions  in higher dimensions.

In addition to applications in quantum field theory,  non-invertible topological defects have also been realized on the lattice \cite{Feiguin:2006ydp,Buican:2017rxc,Aasen:2016dop,Aasen:2020jwb,Inamura:2021szw,Koide:2021zxj,Huang:2021nvb,Vanhove:2021zop}. 
In recent years, this has been described in some condensed matter literature under the name ``algebraic higher symmetry'' \cite{Ji:2019jhk,Kong:2020cie}. 

\begin{center}
	\emph{Duality Defects in Diverse Dimensions}
\end{center}

In this paper, we present a general construction of non-invertible, codimension-one topological defects in even spacetime dimensions $d$ via gauging, both in continuum quantum field theory and on the lattice. 
Our starting point is a $d$-dimensional quantum system ${\cal T}$ with an anomaly-free, abelian $q$-form global symmetry $G^{(q)}$. 
We further assume that ${\cal T}$ is self-dual under gauging:
\ie
{\cal T} \cong {\cal T} /G^{(q)}\,.
\fe
(The more precise meaning of this isomorphism will be elaborated in the main text. See \eqref{selfdual}.) 
Notice that gauging $G^{(q)}$ gives rise to a dual $(d-q-2)$-form symmetry \cite{Gaiotto:2014kfa,Tachikawa:2017gyf}, and hence self-duality is only possible if  $q=(d-2)/2$.

For this class of systems, we can construct a non-invertible topological defect as follows.
We divide the spacetime into two halves,  $L$ and $R$,  and gauge the $G^{(q)}$ symmetry only in $R$ but not in $L$. 
At the codimension-one interface between $L$ and $R$, we impose the topological Dirichlet boundary condition for the discrete $G^{(q)}$ gauge fields.  This then defines a topological defect $\cal D$ in ${\cal T}$.

This general construction encompasses a large class of non-invertible defects. 
When $d=2$, $q=0$ and $G^{(0)}=\mathbb{Z}_2^{(0)}$, this construction gives the Kramers-Wannier duality defect line with the Ising fusion rule:
\ie\label{intro2dfusion}
1+1d:~~~
&\eta\times {\cal D} = {\cal D}\times \eta = {\cal D}\,,~~~~\eta\times \eta=1\,,\\
&{\cal D} \times {\cal D} = 1+\eta\,,
\fe
where $\eta$ is the $\mathbb{Z}_2^{(0)}$ symmetry line. 
We will therefore refer to this general class of topological defects $\cal D$ from gauging as the \textit{duality defects}.  
In particular, the duality defect line ${\cal D}$ is non-invertible, i.e. it does not have an inverse  such that the product with itself is the identity line.

The primary focus of this paper is the case of $d=4$ and $q=1$.  
We derive the fusion rule of the 3-dimensional duality defect $\cal D$ and the one-form symmetry surfaces $\eta$ of $ G^{(1)}$: 
\ie\label{introfusion}
3+1d:~~~
&\eta\times {\cal D}= {\cal D}\times \eta= {\cal D}\,,\\
&{\cal D} \times \overline{\cal D} = {1\over |G|} \sum_{S\in H_2(M;G)}\eta
(S)\,.
\fe
Here $M$ is a connected, orientable 3-manifold on which the duality defect is supported, 
 and $\overline{\cal D}$ is the orientation-reversal of $\cal D$.\footnote{In 1+1d, the duality defect in \eqref{intro2dfusion} is its own orientation-reversal, i.e. ${\cal D}=\overline{\cal D}$ \cite{TAMBARA1998692}. See Section \ref{sec:interface} for more discussions.} 
The fusion  ${\cal D}\times \overline{\cal D}$ is  a special case of the ``condensation"  in \cite{Gaiotto:2019xmp}.

Recently, the authors of \cite{Koide:2021zxj}, based on earlier work of \cite{Aasen:2016dop}, constructed an interesting 3+1d lattice model of $\mathbb{Z}_2$ gauge theory that exhibits a non-invertible defect that implements the Kramers-Wannier-Wegner duality \cite{Wegner:1971app}.   
Various general properties, such as the quantum dimension and certain crossing relations, have been demonstrated from that lattice model. 
In the current paper, we provide a complementary construction of this duality defect from gauging one-form symmetry. 
In particular, we will discuss an alternative  lattice model for the $\mathbb{Z}_2$ gauge theory that realizes the duality defect. 

\begin{center}
	\emph{Dynamical Consequences}
\end{center}

The existence of the duality defect $\cal D$ in a 3+1d system ${\cal T}$ has important consequences for renormalization group  flows.

We derive general constraints by analyzing the compatibility between a trivially gapped phase and the duality defect. 
The construction of the duality defect employs discrete gauging of the one-form symmetry $G^{(1)}$. 
Since discrete gauging is a topological manipulation, it commutes with the RG flow. 
Therefore, for any flow that preserves this duality defect and $G^{(1)}$, we can construct the defect both in the UV and in the IR from gauging the one-form symmetry $G^{(1)}$. 
It follows that the system in the deep IR must also be self-dual under gauging the one-form symmetry $G^{(1)}$, i.e.  
\ie\label{IRselfdual}
{\cal T}_{\text{IR}} \cong {\cal T}_{\text{IR}} /G^{(1)}\,.
\fe

The condition \eqref{IRselfdual} places stringent constraints on the IR phase of the system. 
Suppose the low energy phase is trivially gapped with a unique ground state with no long-range topological order. 
When we activate the background gauge fields $A^{(2)}$ for $G^{(1)}$, this gapped phase is described by a symmetry protected topological (SPT) phase for the one-form global symmetry $G^{(1)}$. 
However, for some $G^{(1)}$, we show that there is no $G^{(1)}$-SPT that obeys \eqref{IRselfdual}. 
Specifically, when $G^{(1)}=\mathbb{Z}_N^{(1)}$, we prove:

\paragraph{Theorem} \emph{Let $\mathcal{T}$ be any 3+1d bosonic QFT which is invariant under gauging a $\mathbb{Z}_{N}^{(1)}$ one-form symmetry.  Then $\mathcal{T}$ can flow to a trivially gapped phase only if each prime factor of $N$ is one modulo four.}

Similarly, in the case of fermionic theories we derive:

\paragraph{Theorem} \emph{Let $\mathcal{T}$ be any 3+1d fermionic QFT which is invariant under gauging a $\mathbb{Z}_{N}^{(1)}$ one-form symmetry.  Then $\mathcal{T}$ can flow to a trivially gapped phase only if $N$ is even and each prime factor of $N/2$ is one modulo four.}

These theorems imply that the existence of certain duality defects is intrinsically incompatible with a trivially gapped phase. This is analogous to constraints implied by 't Hooft anomalies, and generalizes the results of \cite{Chang:2018iay} in 1+1d.

\begin{center}
	\emph{Examples of Duality Defects in 3+1 Dimensions}
\end{center}

The simplest realization of a duality defect in 3+1 dimensions is in free $U(1)$ gauge theory, which has an electric $U(1)_e^{(1)}$ one-form global symmetry. 
Gauging a $\mathbb{Z}_N^{(1)}$ subgroup changes the complexified coupling $\tau$ to $\tau/N^2$. 
Using the S-duality $\tau\to -1/\tau$, we note that the $U(1)$ gauge theory is self-dual under gauging $\mathbb{Z}_N^{(1)}$ at
\ie
\tau = i N \,.
\fe

The continuum Lagrangian for the duality defect in free $U(1)$ gauge theory at $\tau=iN$ is:
\ie
S = {N\over 4\pi} \int_{x<0} d A_L\wedge \star dA_L
+{N\over 4\pi} \int_{x>0} d A_R\wedge \star dA_R
+{iN\over 2\pi}\int_{x=0} A_L \wedge dA_R
\fe
where $A_L$ and $A_R$ are the dynamical one-form gauge fields in region $L:x<0$ and region $R:x>0$. 
The duality defect, supported on the interface $M:x=0$, is realized as an off-diagonal Chern-Simons coupling between the gauge fields on the two sides. 
We also give a lattice realization of the duality defect, using a modified Villain lattice model of the 3+1d $U(1)$ gauge theory that has been developed in \cite{Sulejmanpasic:2019ytl,Gorantla:2021svj}.

As another example, we realize the duality defect in the 3+1d $\mathbb{Z}_N$ lattice gauge theory in the Villain formulation.  
In the case of $\mathbb{Z}_2$, this can be viewed as a complementary lattice realization of the duality defect to the work of \cite{Koide:2021zxj}.  
At the self-dual point, the duality defect is realized as a Chern-Simons coupling between the $\mathbb{Z}_N$ gauge fields from the two sides:
\ie
\frac{\pi}{N} \sum_{\text{plaquette}}\left(\Delta m_L^{(1)}-Nn_L^{(2)}\right)^2
+\frac{\pi}{N} \sum_{\text{plaquette}}\left(\Delta m_R^{(1)}-Nn_R^{(2)}\right)^2
+\frac{2\pi i}{N}\sum_{\text{defect}}m_L^{(1)}\Delta  m_R^{(1)}\,.
\fe
where $m^{(1)}, n^{(2)}$ are $\mathbb{Z}_N$ one- and two-form gauge fields and $L,R$ denote which side of the duality defect they reside in.  
We apply our general theorem  from the SPT analysis to the low-energy limit of the $\mathbb{Z}_2$ lattice gauge theory, and find that it is consistent with the expected first-order phase transition.

We further identify a duality defect from gauging a $\mathbb{Z}_2^{(1)}\times \mathbb{Z}_2^{(1)}$ one-form symmetry in the 3+1d $SO(8)$ gauge theory, and more generally in some gauge theories theories with Lie algebra $\frak{spin}(4n)$.

In addition to the above 3+1d examples, we also realize the duality line in the 1+1d compact boson CFT at special radii, both in the continuum and in its modified Villain lattice model. 
(See also \cite{Ji:2019ugf} for the duality lines in the compact boson CFT.)  
The duality line is realized as a 0+1d Chern-Simons coupling in  in the continuum:
\ie
S= {N\over 4\pi} \int d\tau \int_{-\infty}^0 dx (\partial_\mu\phi_L)^2
+{N\over 4\pi} \int d\tau \int_{0}^\infty dx (\partial_\mu\phi_R)^2
+{iN\over2\pi} \int d\tau\phi_L \partial_\tau\phi_R \Big|_{x=0}\,,
\fe
where $\phi_{L}, \phi_R$ are the boson fields in region $L:x<0$ and region $R:x>0$, respectively.

Looking to the future, the idea of non-invertible duality defects from gauging higher form symmetries may be viewed as a special case of interfaces constructed by coupling QFTs to general TQFTs in half of spacetime.  In this broader context, one may utilize the full machinery of TQFTs, e.g.\ the cobordism hypothesis, to exhibit novel non-invertible defects in QFTs. (See in particular \cite{Freed:2018cec} for a related mathematical discussion on Kramers-Wannier duality.\footnote{We thank D. Freed, and C. Teleman for several enlightening discussions on this topic.})  Clearly, this is just the beginning of an exploration of these generalized notions of symmetry and their dynamical implications.

Note Added: The submission of this paper is coordinated with \cite{Kaidi:2021xfk}, where complementary examples of  non-invertible duality defects in 3+1d are explored.

\section{Non-invertible Topological Defects from Gauging}\label{gaugingsec}

In this section we present a general discussion of topological interfaces and defects that result from gauging higher-form global symmetry in half of the spacetime. 
This construction encompasses the Kramers-Wannier duality defect line in 1+1d as well as the 3-dimensional duality defect in 3+1d.

\subsection{Topological Interfaces and Defects from Gauging}\label{sec:interface}

Consider  a quantum system ${\cal T}^{(q)}$ on a closed $d$-dimensional  spacetime manifold $X$ with an abelian,  discrete $q$-form global symmetry $G^{(q)}$.  
We will assume $G^{(q)}$ to be free of 't Hooft anomaly, i.e.\ it can be gauged.\footnote{In general we only require that anomalies which obstruct gauging $G^{(q)}$ vanish.  Mixed anomalies involving $G^{(q)}$ and other backgrounds need not vanish. } 
For concreteness, we will assume $G^{(q)}=\mathbb{Z}_N^{(q)}$, but the discussion directly generalizes to any abelian finite group.

\begin{figure}
    \centering
    \includegraphics[width=0.5\textwidth]{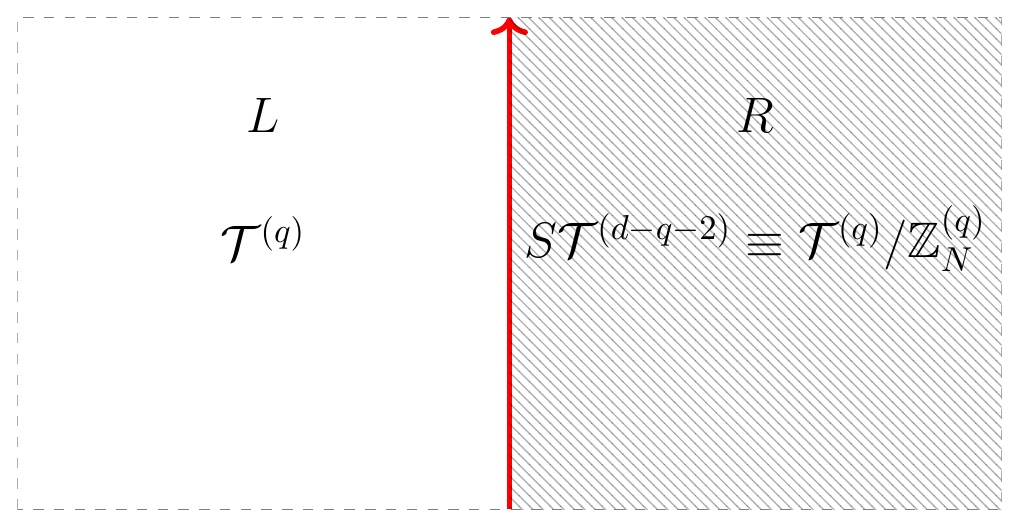}
    \caption{The spacetime manifold is divided into two regions along the interface which is depicted as a red line.  The arrow on the interface indicates that the interface is generally oriented.}
    \label{fig:1}
\end{figure}

Next, we divide the spacetime into two regions, $L$ and $R$, and let $M$ be the codimension-one interface between them.  
See Figure \ref{fig:1}.  

Now, we will gauge the $\mathbb{Z}_N^{(q)}$ global symmetry in half of the spacetime $R$. 
To do this, we  couple the system ${\cal T}^{(q)}$ in $R$ to a dynamical $\mathbb{Z}_N^{(q)}$  gauge theory with flat $(q+1)$-form discrete gauge fields $a^{(q+1)}$.  
More specifically, the $\mathbb{Z}_N^{(q)}$ gauge theory can be represented by the Lagrangian \cite{Maldacena:2001ss,Banks:2010zn,Kapustin:2014gua}:
\ie\label{ZNtheory}
{i N\over 2\pi} b^{(d-q-2)} da^{(q+1)}
\fe
where the superscripts indicate the form degree of the gauge fields. 
Here $b^{(d-q-2)}$ is a $(d-q-2)$-form $U(1)$ gauge fields which serves as a Lagrange multiplier enforcing $a^{(q+1)}$ to be $\mathbb{Z}_N$-valued.
At the interface $M$ between $L$ and $R$, we impose the Dirichlet boundary condition:
\ie\label{Dbc}
a^{(q+1)} \Big|_{M} = 0
\fe
where $(\cdot)|_{M}$ means the restriction of the $(q+1)$-form on the interface $M$. 
The full action is then
\ie
\int_{L} {\cal L}_{{\cal T}^{(q)}}
+\int_R {\cal L}_{{\cal T}^{(q)}} [a^{(q+1)}]
+{iN\over 2\pi} \int_R  b^{(d-q-2)} da^{(q+1)}
\fe
where ${\cal L}_{{\cal T}^{(q)}} [a^{(q+1)}]$ is the Lagrangian of the system ${\cal T}^{(q)}$ coupled to the gauge field $a^{(q+1)}$. 

In  region $L$, we have the original system ${\cal T}^{(q)}$ with a $\mathbb{Z}_N^{(q)}$ $q$-form global symmetry. 
In region $R$, we have the gauged system 
\ie
S{\cal T}^{(d-q-2)} \equiv {\cal T}^{(q)}/\mathbb{Z}_N^{(q)}\,.
\fe
Here we denote the gauging by $S$ and the gauged system by $S{\cal T}^{(d-q-2)}$. 
The latter has a dual $\mathbb{Z}_N^{(d-q-2)}$ $(d-q-2)$-form global symmetry \cite{Gaiotto:2014kfa,Tachikawa:2017gyf}, which is the generalization of the quantum symmetry for orbifolds in 1+1d \cite{Vafa:1989ih}.  
The dual symmetry is generated by the topological operator 
\ie
\eta\equiv \exp(  {i }  \oint a^{(q+1)})\,,
\fe
 with $\eta^N=1$.

In the pure $\mathbb{Z}_N^{(q)}$ gauge theory \eqref{ZNtheory}, the Dirichlet boundary condition $a^{(q+1)}|=0$ is topological. 
To see this, if we deform the locus of the boundary slightly, the difference is computed by $da^{(q+1)}$, which is zero because of   the flatness condition of $a^{(q+1)}$. 
After we couple the gauge theory to ${\cal T}^{(q)}$, the gauge field $a^{(q+1)}$ is still flat from the equation of motion of $b^{(d-q-2)}$. 
Hence the Dirichlet boundary condition \eqref{Dbc} defines a \textit{topological interface} ${\cal D}$  between the two systems, ${\cal T}^{(q)}$ and $S {\cal T}^{(d-q-2)}$.
See Section  \ref{sec:latticetop} for a detailed demonstration of the topological nature of the interface in the lattice model examples.

Generally, the systems ${\cal T}^{(q)}$ and $S {\cal T}^{(d-q-2)}$ are distinct.  
We will be particularly interested in the case when they are isomorphic (dual) to each other, 
\ie
{\cal T}^{(q)}  \cong S{\cal T}^{(d-q-2)} \equiv {\cal T}^{(q)}/\mathbb{Z}_N^{(q)}\,.
\fe
This can only be the case when 
\ie\label{qd}
q = {d-2\over2}\,.
\fe
For example, this includes $\mathbb{Z}_N^{(0)}$ in $d=2$, $\mathbb{Z}_N^{(1)}$ in $d=4$, and $\mathbb{Z}_N^{(2)}$ in $d=6$.
In this case, ${\cal D}$ becomes a codimension-one topological defect in that system.  
 We will see that it is in fact a non-invertible defect.

More precisely, we require that the partition functions on any closed $d$-manifold $X$ coupled to the background $(q+1)$-form $\mathbb{Z}_N$ gauge fields $A^{(q+1)}$ is left invariant under gauging:
\ie\label{selfdual}
{\cal Z}_{{\cal T}^{(q)}}[A^{(q+1)} ] =&  { |H^{q-1}(X;\mathbb{Z}_N) ||H^{q-3}(X;\mathbb{Z}_N) |\cdots\over |H^{q}(X;\mathbb{Z}_N) ||H^{q-2}(X;\mathbb{Z}_N) |\cdots}\\
&\times\sum_{a\in H^{q+1} (X;\mathbb{Z}_N)}  {\cal Z}_{{\cal T}^{(q)}}[a^{(q+1)}] \exp\left( {2\pi i\over N}  \int_X a^{(q+1)} \cup A^{(q+1)}\right)\,.
\fe
The above equality is up to $d$-dimensional local counterterms that are independent of $A^{(q+1)}$. 
The factor in front of the summation is the standard normalization factor for a $(q+1)$-form $\mathbb{Z}_N^{(q)}$ gauge field.

Let us discuss the orbit of gauging $S$ in the case when \eqref{qd} is obeyed. 
For a more general discussion, see \cite{Gaiotto:2014kfa,Bhardwaj:2020ymp,Hsin:2021qiy}. 
For simplicity, we will assume the manifold to be orientable. 
For even $q$ (such as $q=0, d=2$ and $q=2, d=6$), starting from a general system ${\cal T}^{(q)}$ that is not necessarily self-dual, gauging twice gives 
\ie
\text{Even } q:~~~{\cal Z}_{S^2{\cal T}^{(q)}}[A^{(q+1)}] = {\cal Z}_{{\cal T}^{(q)}}[A^{(q+1)}]\,,
\fe
up to a Euler counterterm. In other words, $S^2=1$. 
This implies that the duality defect $\cal D$ does not carry an orientation in this case, i.e. ${\cal D}=\overline{\cal D}$. 
On the other hand, when $q$ is odd (such as $q=1, d=4$):
\ie
\text{Odd }q:~~~{\cal Z}_{S^2{\cal T}^{(q)}}[A^{(q+1)}] = {\cal Z}_{{\cal T}^{(q)}}[-A^{(q+1)}]\,.
\fe
In other words, $S^2=C$, where $C$ flips the sign of the background gauge field. 
In particular,  in 3+1d, the duality defect $\cal D$ is distinct from it orientation-reversal $\overline{\cal D}$ unless $N=2$.   
If the theory ${\cal T}^{(1)}$ is self-dual in 3+1d in the sense of \eqref{selfdual}, that further implies that ${\cal Z}_{{\cal T}^{(1)}}[A^{(2)}]={\cal Z}_{{\cal T}^{(1)}}[-A^{(2)}]$.

More generally, in \eqref{selfdual}, we can include a nontrivial SPT term for the gauge fields $a^{(q+1)}$ in the sum. 
We can also use a more general pairing between $a^{(q+1)}$ and $A^{(q+1)}$. 
These variations will lead to more general kinds of non-invertible defects. 
For example, in 1+1d, the duality line and the symmetry lines form a fusion category known as the Tambara-Yamagami category, which depends on the choice of a symmetric, non-degenerate bicharacter that pairs the dynamical and the background gauge fields \cite{TAMBARA1998692}. 
We leave these generalizations for future investigation.

 \subsection{Kramers-Wannier Duality Defect in 1+1d}

 Let us start with the case $d=2$ and $q=0$. 
In this case,  the system ${\cal T}$ has an ordinary, anomaly-free $\mathbb{Z}_N^{(0)}$ global symmetry. 
We assume that the orbifold theory ${\cal T}/\mathbb{Z}_N^{(0)}$ is isomorphic to itself:\footnote{For a non-spin (bosonic) system, there is no nontrivial SPT term for the $\mathbb{Z}_N^{(0)}$ symmetry in 1+1d because $H^2(\mathbb{Z}_N;U(1))=0$.  For more general abelian group $G$, one can add a discrete torsion $H^2(G;U(1))$ when gauging \eqref{selfdual}. }
\ie
{\cal T} \cong {\cal T}/\mathbb{Z}_N^{(0)}\,.
\fe
We will assume that ${\cal T}$ has a unique topological local operator, i.e. the identity operator, that is invariant under the $\mathbb{Z}_N^{(0)}$ symmetry.  

For example, when $N=2$, we can choose ${\cal T}$ to be the critical Ising CFT, and the topological defect is the Kramers-Wannier duality line \cite{Frohlich:2004ef,Frohlich:2006ch,Frohlich:2009gb,Chang:2018iay,Ji:2019ugf}.\footnote{Kramers-Wannier duality states that the Ising model at high temperature is dual to that at low temperature coupled to a $\mathbb{Z}_2^{(0)}$ gauge field. (See, for example, \cite{Kapustin:2014gua,Gorantla:2021svj}, for recent discussions.) At the critical point, this reduces to the statement that the $\mathbb{Z}_2^{(0)}$ orbifold of the critical Ising CFT is itself.}  
For more general $N$, ${\cal T}$ can be taken to be the coset CFT $SU(2)_N/U(1)$.
 
Let us determine the fusion algebra of the duality defect $\cal D$ and the $\mathbb{Z}_N^{(0)}$ symmetry line $\eta$. 
Since  $\cal D$ is defined as the Dirichlet boundary condition of the gauge field $a^{(1)}| = 0$, the $\mathbb{Z}_N^{(0)}$ line $\eta = \exp({i } \oint a^{(1)})$  vanishes on the defect. We therefore conclude ${\cal D}\times\eta =\eta\times {\cal D} ={\cal D}$.

Next, we consider the fusion between ${\cal D}$ and itself. From ${\cal D}\times\eta =\eta\times {\cal D} ={\cal D}$, we know that the only consistent fusion rule is 
\ie\label{2dprefusion}
{\cal D} \times {\cal D} =x \sum_{i=0}^{N-1} \eta^i\,.
\fe
Here we already see that the defect $\cal D$ is non-invertible, i.e. there does not exist another line such that its fusion with $\cal D$ is the identity line. 
It remains to determine  $x$, which is related to the quantum dimension of the duality line as $\langle{\cal D}\rangle^2 = Nx$. 
The quantum dimension of a line is defined as the expectation value of $\cal D$ acting on the state $|1\rangle$ corresponding to the identity operator.

\begin{figure}[ht]
    \centering
    \includegraphics[width=0.35\textwidth]{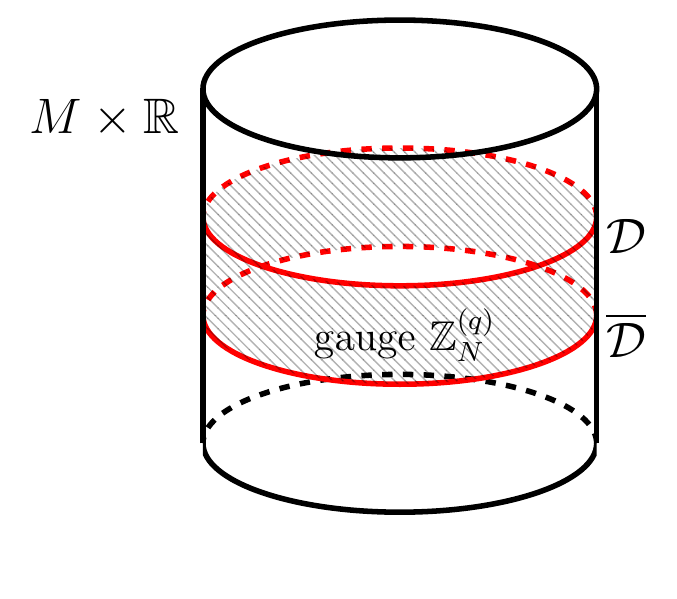}
    \caption{In the shaded region the system is coupled to a $\mathbb{Z}_N^{(q)}$ gauge theory, with Dirichlet boundary conditions imposed on the boundaries. } \label{fig:2}
\end{figure}

Fusion of two such defects on a cylinder  can be understood as follows.
We wrap two duality defects around $M=S^1$ and separate them along $\mathbb{R}$. 
This is equivalent to coupling the system inside the annulus $M\times I$ with a $\mathbb{Z}_N^{(0)}$ gauge theory, with the Dirichlet boundary condition imposed   on the two boundaries. 
See Figure \ref{fig:2}. 
From this picture, the normalization factor $x$ then comes from the standard normalization of a $\mathbb{Z}_N^{(0)}$ gauge theory:
\ie
x= {1\over |H^0 ( S^1\times I , \partial (S^1\times I)  ; \mathbb{Z}_N)|}=1\,.
\fe
Here we use the relative cohomology because of the Dirichlet boundary conditions. It follows that the quantum dimension of the duality defect is:
\ie
\langle {\cal D}\rangle =\sqrt{N}\,,
\fe

To conclude, the fusion algebra of the non-invertible line $\cal D$ and the $\mathbb{Z}_N^{(0)}$ line $\eta$ is
\ie\label{fusion}
&\eta \times {\cal D}  = {\cal D}\times \eta = {\cal D}\,,~~~~~\eta^N=1\,,\\
&{\cal D}\times {\cal D} = \sum_{i=0}^{N-1} \eta^i
\fe
This is indeed the fusion rule of the  Tambara-Yamagami fusion category \cite{TAMBARA1998692}.  
When $N=2$, this reduces to the Ising fusion rule.

\begin{figure}[ht]
    \centering
    \includegraphics[width=0.5\textwidth]{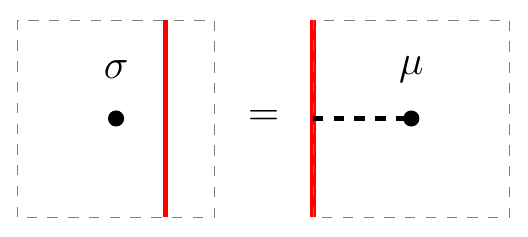}
    \caption{In the 1+1d critical Ising CFT, when the duality line crosses the order operator $\sigma$, it becomes a disorder operator $\mu$ attached to the $\mathbb{Z}_2^{(0)}$ line. Here the red line denotes the duality line $\cal D$ and the dotted line denotes the $\mathbb{Z}_2^{(0)}$ line. }
    \label{fig:3}
\end{figure}

Let us discuss the action of the duality defect on the local operators of ${\cal T}$. 
As we sweep $\cal D$ past a $\mathbb{Z}_N^{(0)}$-charged local operator, it becomes gauged and is now attached to a $\mathbb{Z}_N^{(0)}$ (Wilson) line which intersects with $\cal D$. 
For example, in the critical Ising CFT, as we bring the duality line past the order operator $\sigma$, the latter turns into a disorder operator $\mu$ attached to the $\mathbb{Z}_2^{(0)}$ Wilson line \cite{Frohlich:2004ef}. See Figure \ref{fig:3}.

\subsection{Duality Defect in 3+1d}\label{4dsec}

We now move on to a 3+1d system ${\cal T}$ with an anomaly-free $\mathbb{Z}_N^{(1)}$ one-form global symmetry, which is invariant under gauging $\mathbb{Z}_N^{(1)}$ in the sense of \eqref{selfdual}:
\ie
{\cal T} \cong {\cal T}/\mathbb{Z}_N^{(1)}\,.
\fe
More generally, we can include a nontrivial SPT terms for the gauge field $a^{(2)}$  on the righthand side of  \eqref{selfdual}. 
This will define a different kind of duality defect that we will leave for future studies.  
We will assume ${\cal T}$ has a unique topological local operator, i.e.\ the identity, so that the ground state on $S^3$  is non-degenerate. 

The topological defect $\cal D$ in this case is three-dimensional   and the  $\mathbb{Z}_N^{(1)}$ one-form symmetry defects  $\eta$ are two-dimensional surfaces in spacetime.  

Again since the defect $\cal D$ is defined as the Dirichlet boundary condition $a^{(2)}|=0$, we find that the $\mathbb{Z}_N^{(1)}$ surfaces are annihilated when brought to $\cal D$.

Similar to the 1+1d case, fusion of $\cal D$ and $\overline{\cal D}$ supported on an arbitrary closed 3-surface $M$ can be understood as follows.
The 4-dimensional spacetime is taken to be $M\times \mathbb{R}$. 
We wrap ${\cal D}$ and $\overline{\cal D}$ around $M$ and separate them along $\mathbb{R}$. 
This is equivalent to coupling the system inside the slab $M\times I$ with a $\mathbb{Z}_N^{(1)}$ gauge theory, with the Dirichlet boundary condition imposed   on the two boundaries. 
See Figure \ref{fig:2}.

Gauging the discrete $\mathbb{Z}_N^{(1)}$ symmetry is done by summing over different gauge field configurations, which are given by the degree 2 cohomology classes of the slab with $\mathbb{Z}_N$ coefficient.
The Dirichlet boundary condition instructs us to sum over the cohomology classes relative to the boundary of the slab, that is, the elements in $H^2(M \times I , \partial(M \times I); \mathbb{Z}_N )$. 

 For simplicity, we  assume that $M$ is orientable, and apply the Lefschetz duality, which is a generalized version of the Poincar\'e duality, relating the relative cohomology classes to the ordinary homology classes.
This asserts that $H^2(M \times I , \partial(M\times I); \mathbb{Z}_N ) \cong H_2 (M \times I; \mathbb{Z}_N) \cong H_2 (M; \mathbb{Z}_N)$, where the second isomorphism is given by shrinking the interval.
The sum over different gauge field configurations is then equivalently understood as summing over different 2-cycles on $M$ in $H_2 (M; \mathbb{Z}_N)$, and these are precisely the cycles the $\mathbb{Z}_N^{(1)}$ symmetry operators $\eta$ wrap around.
Thus, the fusion rule for $\cal D$ and $\overline{\cal D}$ wrapping around an orientable 3-manifold $M$ is:\ie\label{prefusion}
{\cal D}\times \overline{\cal D} 
=  x \sum_{S\in H_2(M ; \mathbb{Z}_N) } \eta (S)  \,,
\fe
where $x$ is a normalization coefficient coming from gauging in the slab.
 
The normalization $x$ is given by the standard normalization for the $\mathbb{Z}_N^{(1)}$ gauge theory:
\ie
x = {|H^0(M\times I ,\partial (M\times I);\mathbb{Z}_N)|\over |H^1(M\times I , \partial (M\times I);\mathbb{Z}_N)|} = {1 \over N} \,.
\fe
Again, this normalization factor is expressed in terms of the relative cohomology groups because of the Dirichlet boundary conditions.\footnote{For any arbitrary manifold $M$  (orientable or not), and for any abelian group $G$, we have $H^*(M\times I, \partial(M\times I );G) \cong \tilde{H}^*((M \times I)/\partial(M \times I );G) \cong \tilde{H}^*(SM \vee S^1;G) \cong \tilde{H}^{*-1} (M;G) \oplus \tilde{H}^{*}(S^1;G) \cong H^{*-1}(M;G)$ at all degrees. Here $SM$ is the suspension of $M$, and the tilde denotes the reduced cohomology groups.
An intuitive way to understand this is that any $n$-cycle in $M$ becomes a relative $(n+1)$-cycle stretching between the boundaries in $M \times I$, which may be immediately seen from the cellular homology.
This then gives us $x = 1/|G|$ for any connected $M$.}

To summarize, the fusion rule in 3+1d is
\ie
&\eta\times {\cal D}={\cal D}\times \eta={\cal D}\,,\\
&{\cal D}\times \overline{\cal D} 
=  {1\over N} \sum_{S\in H_2(M  ; \mathbb{Z}_N) } \eta (S)  \,,
\fe
where $M$ is a connected, orientable 3-manifold on which the duality defects are supported.  
The right-hand side can be viewed as a ``condensation" of the one-form symmetry defects on the worldvolume of the duality defect $M$ \cite{Gaiotto:2019xmp}.

The quantum dimension of the duality defect $\cal D$ on $S^3$ is defined as the eigenvalue of $\cal D$ acting on the state $|1\rangle$ on $S^3$  corresponding to the identity operator. 
From \eqref{fusion}, this gives
\ie\label{qdimS3}
\langle {\cal D}\rangle_{S^3}  = {1\over \sqrt{N}}\,,
\fe
which is consistent with the result in \cite{Koide:2021zxj} when $N=2$. 
The ground states on a more general 3-manifold $M$ will depend on the quantum system ${\cal T}$, and are generally degenerate. 
Hence  there is no canonical, model-independent way to define the quantum dimension of $\cal D$ on a general 3-manifold other than $S^3$.

\begin{figure}[ht]
    \centering
    \includegraphics[width=0.5\textwidth]{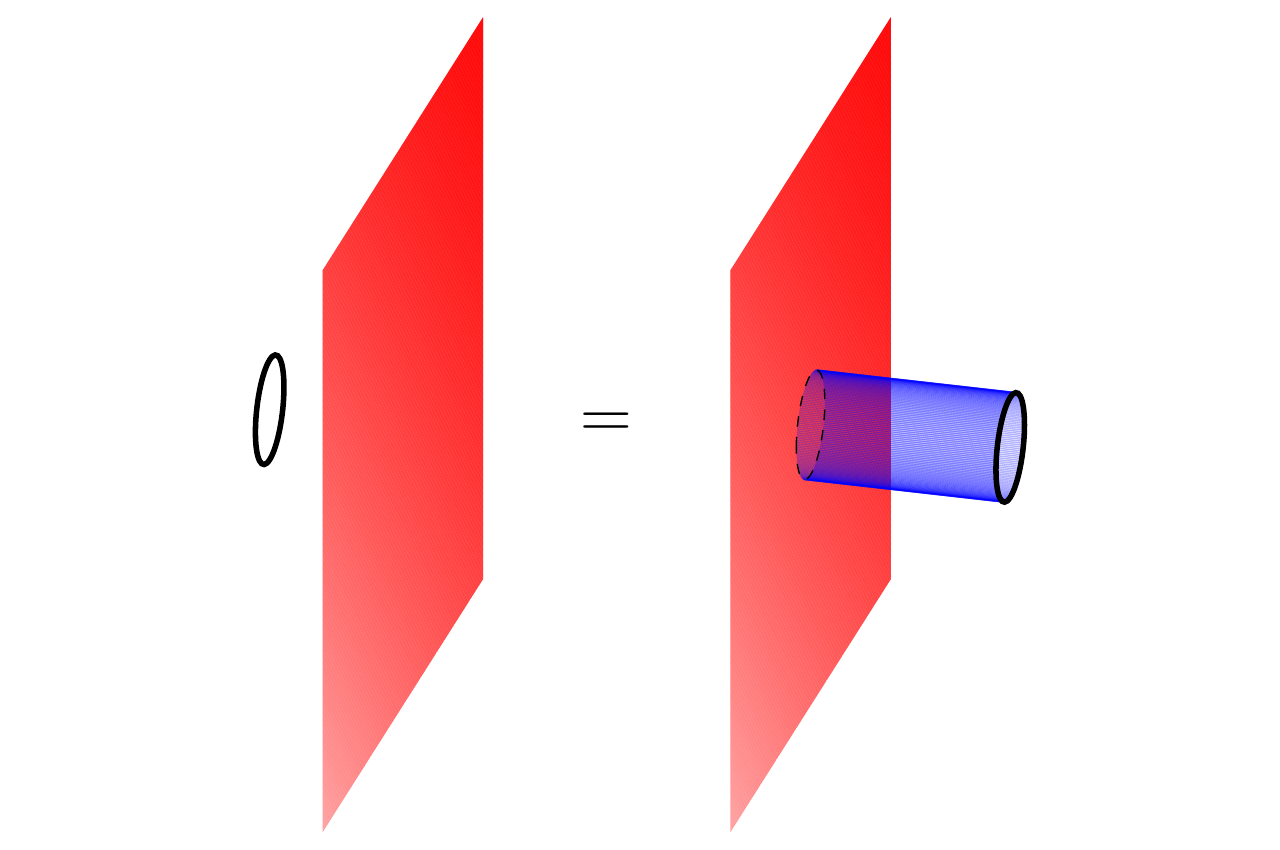}
    \caption{As the three-dimensional duality defect $\cal D$ (shown in red) sweeps past a $\mathbb{Z}_N^{(1)}$-charged line operators  (shown in black), the latter is attached to the $\mathbb{Z}_N^{(1)}$ topological two-surface $\eta$ (shown in blue) on the other side. }
    \label{fig:4}
\end{figure}

Let us discuss the action of the duality defect on the line operators of ${\cal T}$. 
As we sweep $\cal D$ past a $\mathbb{Z}_N^{(1)}$-charged line operator, it becomes gauged and is now attached to a $\mathbb{Z}_N^{(1)}$ surface which intersects with $\cal D$. 
In other words, the duality defect turns a genuine line operator on the one side to a line operator bounded by a topological surface on the other side. 
See Figure \ref{fig:4}. 

\begin{figure}[ht]
    \centering
    \includegraphics[width=0.5\textwidth]{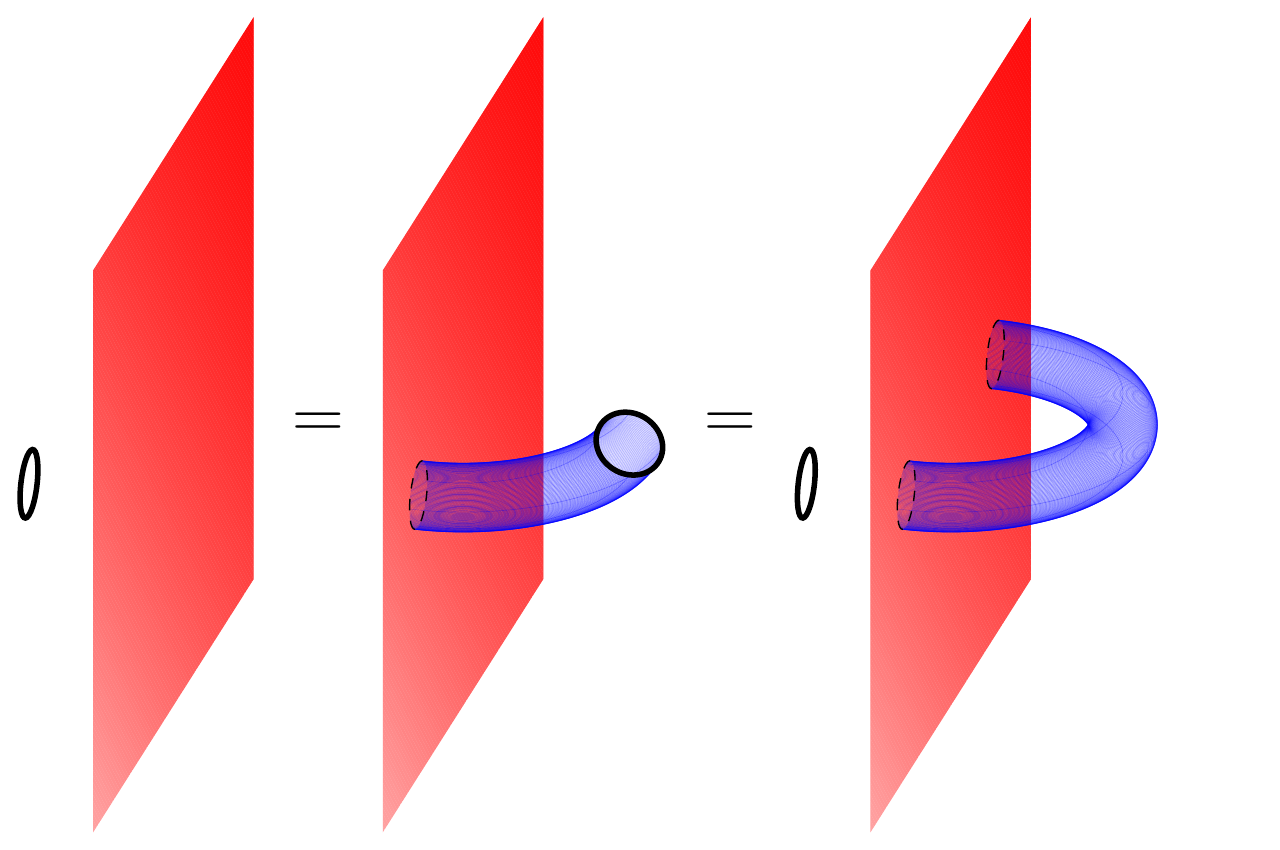}
    \caption{By moving the duality defect $\cal D$ around the charged line operator, the $\mathbb{Z}_N^{(1)}$ topological surfaces $\eta$  can be freely generated and be absorbed by the duality defect, i.e. $\eta\times {\cal D}={\cal D}\times \eta={\cal D}$.} 
    \label{fig:5}
\end{figure}

\section{Dynamical Consequences}\label{dynamicalsec}

The existence of the duality defect  has significant implications for renormalization group flows which we explore in this section.  
We derive  general constraints on renormalization group flows  from a direct analysis of duality defects in trivially gapped phases.\footnote{Mathematically, our results are reminiscent of finding a  fiber functor for a Tambara-Yamagami category in 1+1d \cite{TY} (see also \cite{Thorngren:2019iar,Thorngren:2021yso}).}
We will find that certain kinds of duality defects are intrinsically incompatible with a trivially gapped phases.

We start with a microscopic quantum system $\cal T$ with a  duality defect $\mathcal{D}$ associated with a $\mathbb{Z}_N^{(1)}$ one-form symmetry. 
We assume  that the theory $\mathcal{T}$ is trivially gapped at long distances and will aim to derive a contradiction for some values of $N$.

If the theory $\mathcal{T}$ is trivially gapped, then at long distances, it is described by an SPT (local counterterm) for the one-form global symmetry $\mathbb{Z}_{N}^{(1)}$.  The operation of gauging this symmetry is topological and hence commutes with the renormalization group flow.  Thus, we must be able to reconstruct the duality defect $\mathcal{D}$ by gauging at long distances.  If we denote by $\mathcal{Z}_{\text{SPT}}(A^{(2)})$ the partition function of the SPT, then invariance under gauging implies the equation:
\begin{equation}\label{sptconst}
\mathcal{Z}_{\text{SPT}}(A^{(2)})=\sum_{a^{(2)}\in H^2(X;\mathbb{Z}_N)}\mathcal{Z}_{\text{SPT}}[a^{(2)}]\exp\left(\frac{2\pi i}{N}\int_{X}a^{(2)}\cup A^{(2)}\right)~,
\end{equation}
where the summation above denotes the gauging of the dynamical field $a^{(2)}$ and on the right-hand side $A^{(2)}$ is the dual $\mathbb{Z}_{N}^{(1)}$ background field that arises after gauging.\footnote{As discussed in Section \ref{gaugingsec}, we impose this equality modulo local counterterms that do not depend on the background $A^{(2)}$.}
 Similar SPT analyses in 1+1d were done in \cite{Thorngren:2019iar,Thorngren:2021yso}.

\begin{figure}[ht]
    \centering
    \includegraphics[width=0.4\textwidth]{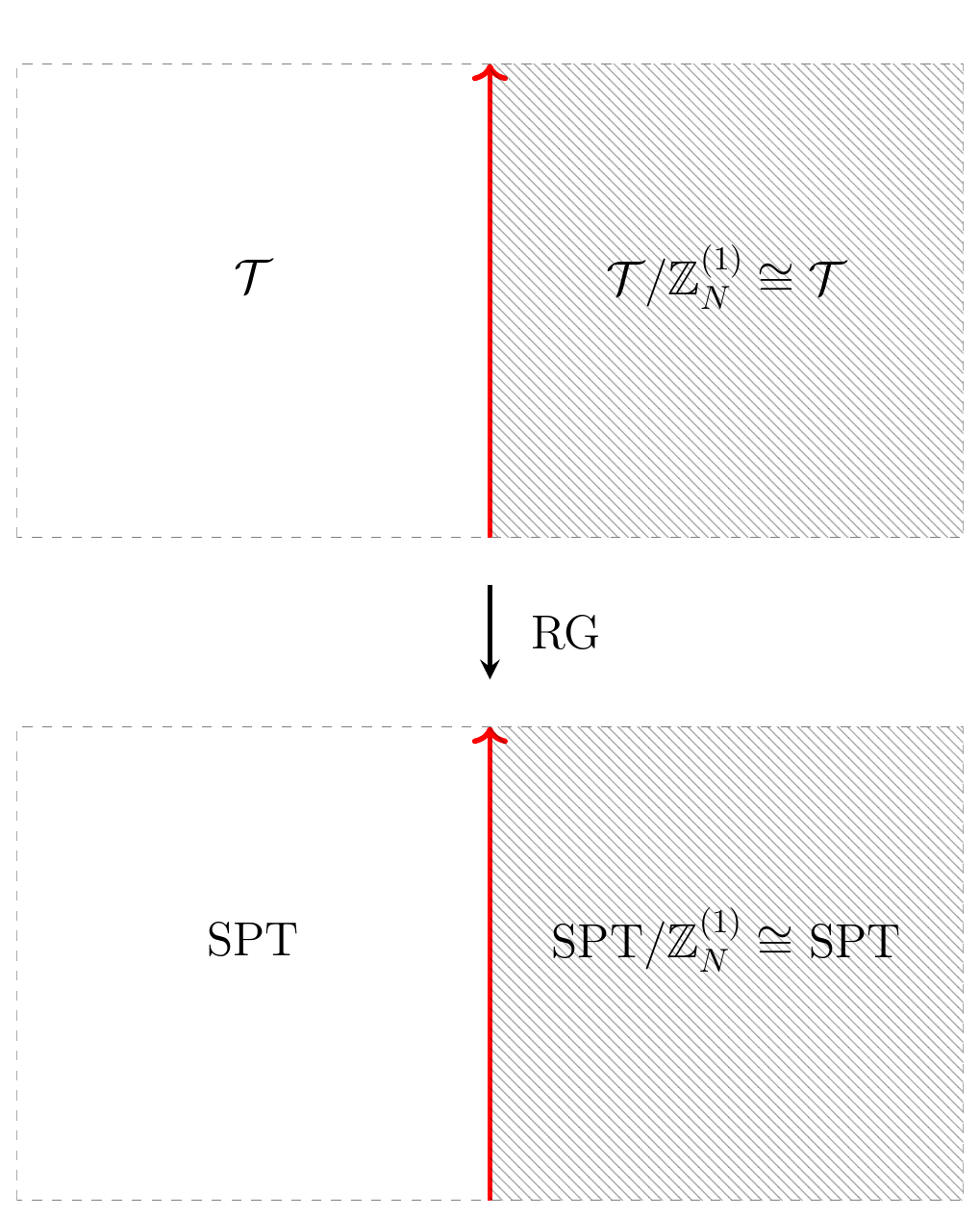}
    \caption{Renormalization group flow commutes with the discrete gauging.  Suppose the low energy phase is trivially gapped. It becomes an SPT phase protected by the one-form global symmetry $\mathbb{Z}_N^{(1)}$ when we activate the background gauge fields for $\mathbb{Z}_N^{(1)}$. 
    This $\mathbb{Z}_N^{(1)}$-SPT phase is compatible with the duality defect if  it is invariant under gauging $\mathbb{Z}_N^{(1)}$.}
    \label{fig:6}
\end{figure}

We now solve this constraint \eqref{sptconst} directly.  
  There are two cases depending on the parity of $N$.

\subsection{SPT Analysis for $N$ Odd}

The most general bosonic  SPT for $\mathbb{Z}_{N}^{(1)}$ in 3+1d is labelled by an integer $p$ defined modulo $N$ and takes the form \cite{Kapustin:2013uxa,Gaiotto:2014kfa,Thorngren:2015gtw,Hsin:2018vcg}:
\begin{equation}\label{bbtheory}
\mathcal{Z}_{\text{SPT}}[A^{(2)}]=\exp\left(\frac{2\pi i p}{N}\int_{X} A^{(2)}\cup A^{(2)}\right)~.
\end{equation}
In general, promoting $A^{(2)}$ to be dynamical results in a non-trivial TQFT.  However, when 
\ie
\gcd(p,N)=1
\fe
 the theory remains trivially gapped even after gauging.  Thus from now on we assume this condition.  

Next, we evaluate the right-hand side of \eqref{sptconst} with the SPT given by \eqref{bbtheory}.  The partition function is
\begin{equation}\label{part1}
\sum_{a^{(2)}}\exp\left(\frac{2\pi i p}{N}\int_{X} a^{(2)}\cup a^{(2)}+\frac{2\pi i }{N}\int_{X} a^{(2)}\cup A^{(2)}\right)~.
\end{equation}
Since the action is quadratic the above can be determined by evaluating on shell.  The equations of motion give:
\begin{equation}
a^{(2)}=-\frac{1}{2p}A^{(2)}~~(\text{mod}~N)~,
\end{equation}
where we have used the fact that $\gcd(2p,N)=1$ to divide by $2p$ in $\mathbb{Z}_{N}$. 
Substituting back into \eqref{part1} we see that the action returns to the original form \eqref{bbtheory} with the transformation $p\mapsto -1/4p$ mod $N$.  Therefore there exists an SPT that is invariant under gauging $\mathbb{Z}_{N}^{(1)}$ if and only if we can find a $p$ such that
\begin{equation}\label{qres}
4p^{2}=-1 ~~(\text{mod}~ N)~.
\end{equation}
In turn, this equation is solvable if and only if $-1$ is a quadratic residue in $\mathbb{Z}_{N}$.  The solution is well known.  One considers the prime factorization of $N$:
\begin{equation}
N=y_{1}^{\ell_{1}}\cdot y_{2}^{\ell_{2}} \cdots y_{m}^{\ell_{m}}~,
\end{equation}
where $y_{i}$ are primes and $\ell_{i}$ positive integers.  
$-1$ is  a quadratic residue if and only if it  is  a quadratic residue separately for each prime $y_{i}$.  
This is true exactly when each prime $y_{i}=1~(\text{mod}~ 4)$.

In summary, for odd integer $N$ there exists an SPT for $\mathbb{Z}_{N}^{(1)}$ which is self-dual under gauging exactly when the factorization of $N$ consists only of primes that are one modulo four.\footnote{Note that this implies that $N$ itself is one modulo four, but it is stronger. For example, $9=1~~(\text{mod}~ 4)$, but its prime factor is not 1 mod 4 so there is no solution to $q^{2}=-1 ~~(\text{mod}~ 9)$.} When this condition on $N$ is satisfied we may thus construct a duality defect in a trivially gapped phase.  In contrast when $N$ does not satisfy this factorization condition, the duality defect cannot be constructed in the putative trivially gapped phase and hence such a phase is excluded.  This establishes the following result.
\paragraph{Theorem} \emph{Let $\mathcal{T}$ be any 3+1d (bosonic or fermionic) QFT which is invariant under gauging a $\mathbb{Z}_{N}^{(1)}$ one-form symmetry, with $N$ odd.  Then $\mathcal{T}$ can flow to a trivially gapped phase only if each prime factor of $N$ is one modulo four.}

If we further assume time-reversal invariance, then there is simply no time-reversal invariant  SPT \eqref{bbtheory} satisfying $\text{gcd}(p,N)=1$.  
Therefore we find that any time-reversal invariant ${\cal T}$ that is invariant under gauging $\mathbb{Z}_N^{(1)}$  cannot flow to a trivially gapped phase.

\subsection{SPT Analysis for $N$ Even}

We can proceed analogously for $N$ even.  In this case the possible bosonic SPTs for $\mathbb{Z}_{N}^{(1)}$ are classified by an integer $p$ defined modulo $2N$, and the partition function takes the form \cite{Kapustin:2013uxa,Gaiotto:2014kfa,Thorngren:2015gtw,Hsin:2018vcg}:
\begin{equation}\label{pbbtheory}
\mathcal{Z}_\text{SPT}[A^{(2)}]=\exp\left(\frac{2\pi i p}{2N}\int_{X} \mathcal{P}(A^{(2)})\right)~,
\end{equation}
where $\mathcal{P}$ denotes the Pontryagin square operation: for a $\mathbb{Z}_N$ two-cocycle $B$, ${\cal P}(B)=B\cup B-B\cup_1 \delta B$. 
 The condition that the theory remains invertible after gauging is again that $\gcd(p,N)=1$.

We can again evaluate the partition function in \eqref{sptconst} by solving the equations of motion of the quadratic action and evaluating the result.  Doing so we deduce that $p$ is transformed as $p\mapsto -1/p$.  Therefore, as before, there exists an SPT invariant under gauging $\mathbb{Z}_{N}^{(1)}$ if and only if $-1$ is a quadratic residue in $\mathbb{Z}_{2N}$,
\begin{equation}
p^{2}=-1 ~~(\text{mod}~ 2N)~.
\end{equation}
Since $N$ is even, we can reduce this equation modulo four and it is then manifest that no such $p$ exists.  Therefore we cannot construct a duality defect in any $\mathbb{Z}_{N}^{(1)}$ SPT and hence we have proven the following result.
\paragraph{Theorem} \emph{Let $\mathcal{T}$ be any 3+1d bosonic QFT which is invariant under gauging a $\mathbb{Z}_{N}^{(1)}$ one-form symmetry, with $N$ even.  Then $\mathcal{T}$ cannot flow to a trivially gapped phase.}

We remark that when we allow the QFT to be fermionic, then $p\sim p+N$, and the condition for the SPT phase to be invariant under gauging the $\mathbb{Z}_N^{(1)}$ one-form symmetry is instead
\begin{equation}
p^{2}=-1 ~~(\text{mod}~ N)~.
\end{equation}
Following a similar discussion in the case of odd $N$, we find:
\paragraph{Theorem} \emph{Let $\mathcal{T}$ be any 3+1d  fermionic QFT which is invariant under gauging a $\mathbb{Z}_{N}^{(1)}$ one-form symmetry, with $N$ even.  Then $\mathcal{T}$ can flow to a trivially gapped phase only if each prime factor of $N/2$ is one modulo four.}

In particular, it is possible for a fermionic QFT that is invariant under gauging $\mathbb{Z}_2^{(1)}$ to flow to a trivially gapped phase. 
This is because there is a fermionic $\mathbb{Z}_2^{(1)}$ SPT (\eqref{pbbtheory} with $N=2$ and $p=1$) that is invariant under gauging $\mathbb{Z}_2^{(1)}$, but it is not invariant when viewed as a bosonic $\mathbb{Z}_2^{(1)}$ SPT.

\section{Continuum Examples}\label{examplesec}

In this section we discuss explicit examples of duality defects  in the continuum.  Each construction below gives rise to an instance of the general, model-independent, fusion rules derived in Section \ref{gaugingsec}. 

\subsection{1+1d Compact Boson}

We start with a warm-up example in 1+1d. 
Consider a free compact boson CFT of $\phi$ in 1+1d.  
The Euclidean action is
\ie
S= {R^2\over 4\pi} \int d\tau dx( \partial_\mu\phi)^2\,,~~~~\phi\sim \phi+2\pi\,.
\fe
T-duality states that the compact boson $\phi$ at radius $R$ is equivalent to the dual boson $\tilde\phi$ at radius $1/R$, with the two related by:
\ie\label{Tdual}
-i R^2 \partial_\mu\phi = \epsilon_{\mu\nu} \partial_\nu \tilde\phi\,,
\fe

The compact boson CFT has a $U(1)^{(0)}_m$ momentum symmetry that shifts $\phi$ by an angle,  and a $U(1)_w^{(0)}$ winding symmetry that shifts $\tilde\phi$ by an angle. 
These two symmetries have a mixed anomaly, but individually they are non-anomalous. 
Gauging the $\mathbb{Z}_N^{(0)}$ subgroup of $U(1)^{(0)}_m$ changes the radius of $\phi$ from $R$ to $R/N$. 
Combining with T-duality, the compact boson CFT at  $R=\sqrt{N}$ is invariant under gauging the momentum $\mathbb{Z}_N^{(0)}$ symmetry. 
We therefore conclude that the compact boson CFT at $R=\sqrt{N}$ has a Kramers-Wannier duality defect line \cite{Ji:2019ugf}. 

This duality line can be described by the following Lagrangian:
\ie\label{contphi}
S= {N\over 4\pi} \int d\tau \int_{-\infty}^0 dx (\partial_\mu\phi_L)^2
+{N\over 4\pi} \int d\tau \int_{0}^\infty dx (\partial_\mu\phi_R)^2
+{iN\over2\pi} \int d\tau\phi_L \partial_\tau\phi_R \Big|_{x=0}\,.
\fe
Here $\phi_{L}, \phi_R$ are the boson fields in region $L:x<0$ and region $R:x>0$, respectively. 
The duality line is realized by  a 0+1d Chern-Simons coupling between the fields from the two sides.

For $N=1$, this Chern-Simons coupling reduces to that for the $T$-duality defect in \cite{Kapustin:2009av}. 
For general $N$, it is a defect corresponding to the composition of $T$-duality  and gauging the $\mathbb{Z}_N^{(0)}$ momentum symmetry. 
In Section \ref{latticesec}, we present a detailed derivation of this Chern-Simons coupling from a modified Villain lattice realization of the compact boson.

Demanding that the boundary terms from taking the variation of the action to vanish, we find that at $x=0$:
\ie
&\partial_\mu\phi_L\Big\vert_{x=0}=i\epsilon_{\mu\nu}\partial_\nu\phi_R\Big\vert_{x=0}~.
\fe
Using T-duality \eqref{Tdual}, we can rewrite the above equality as 
\ie
\partial_\mu\phi_L= {1\over N }\partial_\mu\tilde \phi_R =i\epsilon_{\mu\nu}\partial_\nu\phi_R\,,
\fe
which can be understood as first gauging $\mathbb{Z}_N^{(0)}$   to reduce the radius by $1/N$, and then perform the T-duality.  

\subsection{3+1d $U(1)$ Gauge Theory}

The simplest example of a duality defect in higher dimensions occurs in free 3+1d $U(1)$ gauge theory with dynamical one-form gauge field $A$.\footnote{This is not to be confused with the two-form background $\mathbb{Z}_N^{(1)}$ gauge field $A^{(2)}$ in the general discussion above. We hope this slight abuse of notations will not cause any confusions.}   This model is defined by a complexified gauge coupling:
\begin{equation}
\tau =\frac{4\pi i}{g^{2}}+\frac{\theta}{2\pi}~.
\end{equation}
We will see that for certain special values of the coupling we can construct a duality defect.

This theory has an electric one-form symmetry $U(1)_e^{(1)}$ and a magnetic one-form symmetry  $U(1)_m^{(1)}$, acting respectively on the Wilson lines $e^{i \oint A}$ and the 't Hooft lines $e^{i \oint \tilde A}$.  
Here $\tilde A$ is the dual one-form gauge field. 
  These symmetries have a mixed anomaly, but individually they are anomaly-free and may be gauged.  Let us focus on a $\mathbb{Z}_{N}^{(1)}$ subgroup of the electric one-form symmetry $U(1)_e^{(1)}$.  
Gauging this symmetry replaces the dynamical gauge field $A$ by $A/N$, and the dual gauge field $\tilde A$ by $ N \tilde A$. 
Since the action is quadratic in $A$ this is equivalent to changing the coupling from $\tau$ to $\tau/N^{2}$.

For general $\tau$, the theories before and after gauging are distinct.  
Following the discussion in Section \ref{gaugingsec}, we can define a topological interface $\cal D$ between the theory at $\tau$ in region $L$ and that at $\tau/N^2$ in region $R$. 
As we bring a  minimal Wilson line $\exp(i\oint_C A)$  charged under $\mathbb{Z}_N^{(1)}$ in $L$ past the topological interface,  it acquires a topological surface. 
In terms of the $U(1)$ gauge theory, once it has crossed the interface into region $R$, the line appears fractionally charged and hence must be attached to a flux surface:
\begin{equation}
\exp \left( \frac{i}{N}\oint_{C} A \right)\equiv \exp \left( \frac{i}{N}\int_{D} F \right)~,
\end{equation}
where $D$ is a surface with $\partial D=C$.  Since the charge is now fractional the definition of the operator depends on the topological choice of $D$.  Moreover, the surface operator on the right-hand side of the above is indeed the dual magnetic one-form symmetry defect $\eta$ created after gauging as expected on general grounds.  
See Figure \ref{fig:4}.

Dually, in region $L$ there are fractionally charged 't Hooft lines such as $\exp({i\over N}\oint \tilde A)$ that are bounded by the  electric one-form symmetry surface:
\begin{equation}
\exp \left( \frac{i}{N}\oint_{C} \tilde A \right)\equiv \exp \left( \frac{i}{N}\int_{D} \tilde F \right)~,
\end{equation}
As we bring it past the interface, it is liberated from the surface and becomes a genuine line with minimal magnetic charge $\exp(i \oint \tilde A)$ in region $R$.

For certain special values of $\tau$, the gauge theories on two sides of the interface are equivalent due to electric-magnetic duality.  This occurs when the couplings are related by an $SL(2,\mathbb{Z})$ transformation, i.e. whenever we can find integers $a,b,c,d$ such that:\footnote{For simplicity, we view the gauge theory as a fermionic  quantum field theory defined only on spin manifolds, so that it enjoys the full $SL(2,\mathbb{Z})$ duality.}
\begin{equation}
\frac{a\tau+b}{c\tau+d}=\frac{\tau}{N^{2}}~, \hspace{.5in}ad-bc=1~.
\end{equation}
For instance, an example of the above occurs when $\tau= iN$ and the duality is the $S$-transformation $\tau \rightarrow -1/\tau$, which exchanges $A$ and $\tilde A$.  Thus in such a theory, we obtain a duality defect obeying the rule \eqref{fusion}.\footnote{The charge conjugation symmetry implies that  the partition function of the Maxwell theory obeys ${\cal Z}[B^{(2)}]= {\cal Z}[-B^{(2)}]$, where $B^{(2)}$ is the two-form background gauge field for the electric one-form symmetry. This is consistent with the general consequence of the self-duality \eqref{selfdual}.}

The duality defect at $\tau=iN$ can be realized as follows:
\ie\label{contU1}
S = {N\over 4\pi} \int_{x<0} d A_L\wedge \star dA_L
+{N\over 4\pi} \int_{x>0} d A_R\wedge \star dA_R
+{iN\over 2\pi}\int_{x=0} A_L \wedge dA_R
\fe
where $A_L$ and $A_R$ are the dynamical one-form gauge fields in region $L:x<0$ and region $R:x>0$. 
The duality defect is realized as an off-diagonal Chern-Simons term between the gauge fields from the two sides of the interface $M:x=0$.

For $N=1$, this Chern-Simons coupling reduces to that for the $S$-duality defect in \cite{Gaiotto:2008ak,Kapustin:2009av}. 
For general $N$, it is a defect corresponding to the composition of $S$-duality  and gauging the $\mathbb{Z}_N^{(1)}$ electric symmetry. 
In Section \ref{latticesec}, we present a derivation of this Chern-Simons  coupling from a  lattice realization of the 3+1d $U(1)$ gauge theory.

Let us consider the variation of the continuum action \eqref{contU1}. 
Demanding the variational terms vanish at $x=0$, we find: 
\ie\label{junction}
 dA_L  \Big|_{x=0}= - i\star dA_R\Big|_{x=0}\,.
\fe
The gauge field $A$ and its dual $\tilde A$ are related by 
\ie
\frac{1}{2\pi}d\tilde A=- {2i\over g^2} \star d A = - {iN\over 2\pi}\star d A\,.
\fe 
Therefore, we can rewrite \eqref{junction} as 
\ie
x=0:~~ dA_L  = {1\over N} d\tilde A_R =  - i\star dA_R\,.
 \fe
 These equalities can be interpreted as first performing a $\mathbb{Z}_N^{(1)}$ gauging to rescale the gauge field by $1/N$, and then performing an $S$-duality transformation.

\subsection{$SO(8)$ Gauge Theory}
\label{so8sec}

The pure $Spin(8)$ gauge theory enjoys a triality  that permutes the vector, fundamental spinor and cospinor representations. 
It has $\mathbb{Z}_2^{(1)}\times\mathbb{Z}_2^{(1)}$ center one-form symmetry that are permuted by triality as in Table \ref{tab:Spin8Z2}, where we list different $\mathbb{Z}_2^{(1)}$ subgroups. 
Thus, gauging different $\mathbb{Z}_2^{(1)}$ subgroup one-form symmetries gives dual theories that are related by the triality of $Spin(8)$ gauge theory. 
More specifically, gauging various $\mathbb{Z}_2^{(1)}$ subgroups one-form symmetries gives the $SO(8),Sc(8),Ss(8)$ gauge theories that are dual to each other.
We can further include matter fields in the adjoint representation which is invariant under the triality.

\begin{table}[t]
\centering
\begin{tabular}{|c|ccc|}
\hline
$\mathbb{Z}_2$ generator & vector & spinor & cospinor \\ \hline
(1,1) & $(+1)$ & $(-1)$ & $(-1)$\\
(1,0) & $(-1)$ & $(+1)$ & $(-1)$ \\
(0,1) & $(-1)$ & $(-1)$ & $(+1)$\\ \hline
\end{tabular}
\caption{Charges of various $\mathbb{Z}_2^{(1)}$ subgroups of the center one-form symmetry that acts on the Wilson lines in $Spin(8)$ gauge theory.}\label{tab:Spin8Z2}
\end{table}

Pure $SO(8)$ gauge theory in $d$ spacetime dimension has a $\mathbb{Z}_2^{(1)}$ electric center one-form symmetry that acts on the Wilson line in the vector representation, and a $\mathbb{Z}_2^{(d-3)}$ $(d-3)$-form symmetry that acts on the basic 't Hooft line. The two symmetries do not have anomaly and can be gauged \cite{Cordova:2017vab}.\footnote{If we include the charge conjugation 0-form symmetry, then the three symmetries have a mixed anomaly.}

Let us now concentrate on the $d=4$ case, where the magnetic symmetry of the $SO(8)$ gauge theory is also a one-form symmetry. 
Starting from the $SO(8)$ gauge theory, we can gauge the  magnetic symmetry to arrive at the $Spin(8)$ gauge theory. 
We can then gauge a $\mathbb{Z}_2^{(1)}$ symmetry of the $Spin(8)$ gauge theory to obtain the $Sc(8)$ gauge theory. 
Combining the two steps together, this implies that we can gauge a $\mathbb{Z}_2^{(1)}\times\mathbb{Z}_2^{(1)}$ symmetry of the $SO(8)$ gauge theory to arrive at the $Sc(8)$ gauge theory, which is isomorphic to the former.  
Following the discussion in Section \ref{gaugingsec}, we conclude that the $SO(8)$ gauge theory has a duality defect $\cal D$ for the $\mathbb{Z}_2^{(1)}\times \mathbb{Z}_2^{(1)}$ symmetry. 

However, the presence of such duality defect does not forbid the low energy physics to be trivially gapped. 
To see this, we note that there exists an SPT phase with $\mathbb{Z}_N^{(1)}\times \mathbb{Z}_N^{(1)}$ one-form symmetry for any $N$ (here $N=2$), such that it is invariant under gauging the one-form symmetry. For even $N$, the SPT phase has the partition function
\begin{align}
{\cal Z}_\text{SPT}[B,B']=
\exp\left(2\pi i\frac{N-1}{2N}{\cal P}(B)-2\pi i\frac{N-1}{2N}{\cal P}(B')\right)~,
\end{align}
where $B,B'$ are the background two-form gauge fields of the two $\mathbb{Z}_N^{(1)}$ one-form symmetries.
For odd $N$, we replace ${\cal P}(B),{\cal P}(B')$ by $B\cup B,B'\cup B'$ in the partition function.
Note that this SPT phase is time-reversal invariant.

We remark that the pure $SO(8)$ Yang-Mills theory with $\theta=0$ is believed to flow to pure $\mathbb{Z}_2$ gauge theory at low energies \cite{Gaiotto:2014kfa}.  The deconfined $\mathbb{Z}_2$ line comes from the UV 't Hooft line operator, while the UV Wilson line confines due to monopole condensation. Thus, the $\mathbb{Z}_{2}^{(1)}$ electric one-form symmetry acts trivially in the IR, while the $\mathbb{Z}_2^{(1)}$ magnetic one-form symmetry is spontaneously broken. The low energy theory is invariant under gauging the $\mathbb{Z}_2^{(1)}\times\mathbb{Z}_2^{(1)}$ one-form symmetry: gauging the magnetic one-form symmetry removes the $\mathbb{Z}_2$ gauge theory, while gauging the electric one-form symmetry produces a gauge theory with a dynamical $\mathbb{Z}_2$ two-form, which is equivalent to ordinary $\mathbb{Z}_2$ gauge theory in 3+1d.

\subsubsection{$Sc(4n)$ and $Ss(4n)$ gauge theory}

The discussion can be generalized to $Sc(4n)$ and $Ss(4n)$ gauge theories that are equivalent due to the isomorphism $Sc(4n)\cong Ss(4n)$, since they are obtained from $Spin(4n)$ gauge theory by gauging different $\mathbb{Z}_2$ one-form symmetries that are exchanged by
the $\mathbb{Z}_2$ charge conjugation symmetry in $Spin(4n)$.\footnote{
When $n=1$, $Sc(4)\cong SU(2)\times SO(3)$ and $Ss(4)\cong SO(3)\times SU(2)$, while $SO(4)=\left(SU(2)\times SU(2)\right)/\mathbb{Z}_2$.}

The $Sc(4n)$ gauge theory has $\mathbb{Z}_2^{(1)}$ electric one-form symmetry and $\mathbb{Z}_2^{(1)}$ magnetic one-form symmetry, and they are non-anomalous. After gauging the $\mathbb{Z}_2^{(1)}\times \mathbb{Z}_2^{(1)}$ one-form symmetry, the theory becomes the $Ss(4n)$ gauge theory, which is dual to $Sc(4n)$ and thus the theory is invariant under gauging $\mathbb{Z}_2^{(1)}\times \mathbb{Z}_2^{(1)}$ one-form symmetries.
Following the discussion in Section \ref{gaugingsec}, we conclude that the $SO(8)$ gauge theory has a duality defect $\cal D$ for the $\mathbb{Z}_2^{(1)}\times \mathbb{Z}_2^{(1)}$ symmetry. 

We remark that the pure $Sc(4n),Ss(4n)$ Yang-Mills theories with $\theta=0$ are also believed to flow to pure $\mathbb{Z}_2$ gauge theory at low energy \cite{Gaiotto:2014kfa}.

\section{Lattice Examples}\label{latticesec}

In this section, we discuss non-invertible topological defects in the modified Villain lattice models \cite{Sulejmanpasic:2019ytl,Gorantla:2021svj}. 
We consider three examples:
\begin{itemize}
\item  1+1d XY-model. It can also be viewed as the lattice version of the $c=1$ compact boson CFT.  The relevant symmetry here is the $\mathbb{Z}_N^{(0)}$ subgroup of the $U(1)_m^{(0)}$ 0-form momentum symmetry.
\item   3+1d $U(1)$ lattice gauge theory. The relevant symmetry is the $\mathbb{Z}_N^{(1)}$ subgroup of the  $U(1)_e^{(1)}$ electric one-form symmetry.  
\item 3+1d $\mathbb{Z}_N$ lattice gauge theory. The relevant symmetry is the  $\mathbb{Z}_N^{(1)}$ electric one-form symmetry.
\end{itemize}

One prominent feature of the modified Villain lattice models is that the vortices/monopoles are completely suppressed by the Lagrange multiplier fields. See \cite{Gross:1990ub} for  an earlier, related approach. 
Therefore these lattice models exhibit the same global symmetries, anomalies, and exact dualities as their continuum counterparts. 
In particular, the modified Villain model for the 3+1d $U(1)$ gauge theory realizes the $S$-duality exactly on the lattice, which is crucial in our construction of the duality defect.

\subsection{1+1d XY-model}

We start with the warm-up example of the 1+1d XY-model in its modified Villain form where the vortices are suppressed. 

\subsubsection{Modified Villain Model}

We define the model on a 2-dimensional Euclidean square lattice.
The modified Villain action for the XY-model is given by \cite{Sulejmanpasic:2019ytl,Gorantla:2021svj}:
\ie \label{XY}
\frac{R^2}{4\pi} \sum_{\text{link}} (\Delta \phi^{(0)} - 2\pi n^{(1)})^2 + i\sum_{\text{plaquette}} \tilde{\phi}^{(0)} \Delta n^{(1)}\,.
\fe
$\phi^{(0)}$, $\tilde\phi^{(0)}$ are real-valued fields on the sites and the dual sites, respectively. $n^{(1)}$ is an integer-valued gauge field on the links.
The superscripts indicate the form degrees of various fields.
$\Delta$ is the differential operator on the lattice. It maps a field $\lambda^{(p)}$ on the $p$-cells to a field $\Delta \lambda^{(p)}$ on the $(p+1)$-cell, which is given by the oriented sum of $\lambda^{(p)}$ on the boundary of the $(p+1)$-cell.  

The theory is subject to the following gauge symmetries:
\ie
\phi^{(0)} &\sim \phi^{(0)} + 2\pi k^{(0)}\,, \\
n^{(1)} &\sim n^{(1)} + \Delta k^{(0)}\,, \\
\tilde{\phi}^{(0)} &\sim \tilde{\phi}^{(0)} + 2\pi \tilde{k}^{(0)} \,.
\fe
$k^{(0)}$ and $\tilde{k}^{(0)}$ are integer-valued gauge parameters on the sites and the dual sites, respectively.
The gauge symmetry effectively makes $\phi^{(0)}$ and $\tilde{\phi}^{(0)}$  $2\pi$-periodic.
The Lagrange multiplier filed $\tilde\phi^{(0)}$ sets the vorticity $\Delta n^{(1)}$ to zero.

The theory realizes an exact T-duality that maps $R\leftrightarrow1/R$ and exchanges $\phi^{(0)}$ and $\tilde{\phi}^{(0)}$  \cite{Sulejmanpasic:2019ytl,Gorantla:2021svj}.
This can be seen by applying the Poisson resummation formula 
\ie\label{Possonresummationi}
\sum_n \exp\left[-\frac{R^2}{4\pi} (\theta-2\pi n)^2 + i n \tilde \theta\right] = \frac{1}{R} \sum_{\tilde n} \exp\left[-\frac{1}{4\pi R^2} (\tilde \theta-2\pi \tilde n)^2 - \frac{i\theta}{2\pi}(2\pi \tilde n-\tilde \theta)\right]~,
\fe
to the sum over the integer $n^{(1)}$.

\subsubsection{Gauging a $\mathbb{Z}_N^{(0)}$ Symmetry}

The theory has a $U(1)_m^{(0)}$ 0-form momentum symmetry which shifts $\phi^{(0)}$ by a constant.
One can gauge its $\mathbb{Z}_N^{(0)}$ subgroup by coupling the theory to a dynamical $\mathbb{Z}_N^{(0)}$ gauge theory in the following way:
\begin{equation} \label{gaugedXY}
\frac{R^2}{4\pi} \sum_{\text{link}}  \left(\Delta \phi^{(0)} - 2\pi n^{(1)} - \frac{2\pi}{N} \hat{n}^{(1)}\right)^2 + i\sum_{\text{plaquette}} \tilde{\phi}^{(0)} \left(\Delta n^{(1)} + \frac{1}{N}\Delta \hat{n}^{(1)}\right) + \frac{2\pi i}{N} \sum_{\text{plaquette}} \hat m^{(0)} \Delta \hat{n}^{(1)} \,.
\end{equation}
The third term represents a $\mathbb{Z}_N^{(0)}$  gauge theory in the presentation of \cite{Gorantla:2021svj}.
$\hat{n}^{(1)}$ is an integer-valued field on the links, and
$\hat m^{(0)}$ is an integer-valued field on the dual sites whose equation of motion constrains $\hat{n}^{(1)}$ to be a flat $\mathbb{Z}_N$ gauge field.
We have the modified gauge transformations:
\ie
\phi^{(0)} &\sim \phi^{(0)} + 2\pi k^{(0)} +\frac{2\pi  q^{(0)}}{N}\,, \\
n^{(1)} &\sim n^{(1)} +\Delta k^{(0)}- {\ell}^{(1)}\,, \\
\tilde{\phi}^{(0)} &\sim \tilde{\phi}^{(0)} + 2\pi \tilde{k}^{(0)}\,, \\
\hat{n}^{(1)} &\sim \hat{n}^{(1)} + \Delta q^{(0)} + N{\ell}^{(1)}\,, \\
\hat m^{(0)} &\sim \hat m^{(0)} - \tilde{k}^{(0)} + N \tilde q^{(0)}\,.
\fe
$q^{(0)}$, $\ell^{(1)}$ and $\tilde q^{(0)}$ are integer-valued gauge parameters on the sites, the links and the dual sites, respectively.

Using the gauge parameters $\ell^{(1)},q^{(0)}$, we can set $n^{(1)},\hat m^{(0)}$ to zero. We then redefine $\varphi^{(0)} \equiv N\phi^{(0)}$ and $\tilde{\varphi}^{(0)} \equiv \tilde{\phi}^{(0)}/N$ and obtain
\begin{equation}
\frac{R^2}{4\pi N^2}\sum_{\text{links}}(\Delta \varphi^{(0)} - 2\pi\hat{n}^{(1)})^2 + i \sum_{\text{plaquette}} \tilde{\varphi}^{(0)} \Delta \hat{n}^{(1)}~.
\end{equation}
The new fields $\varphi^{(0)}$ and $\tilde\varphi^{(0)}$ are effectively $2\pi$-periodic due to the residue gauge symmetry
\ie\label{eq:gaugeafterdualize}
\varphi^{(0)} &\sim \varphi^{(0)}  +2\pi q^{(0)}\,, \\
\hat{n}^{(1)} &\sim \hat{n}^{(1)} + \Delta q^{(0)}\,,
\\
\tilde{\varphi}^{(0)} &\sim \tilde{\varphi}^{(0)} + 2\pi \tilde q^{(0)}\,.
\fe
This is the modified Villain action at radius $R/N$. In particular, due to the T-duality, the model is self-dual under the gauging if $R=\sqrt{N}$.

\subsubsection{Topological Nature of the Gauging Interface}\label{sec:latticetop}

\begin{figure}[ht]
    \centering
    \includegraphics[width=0.4\textwidth]{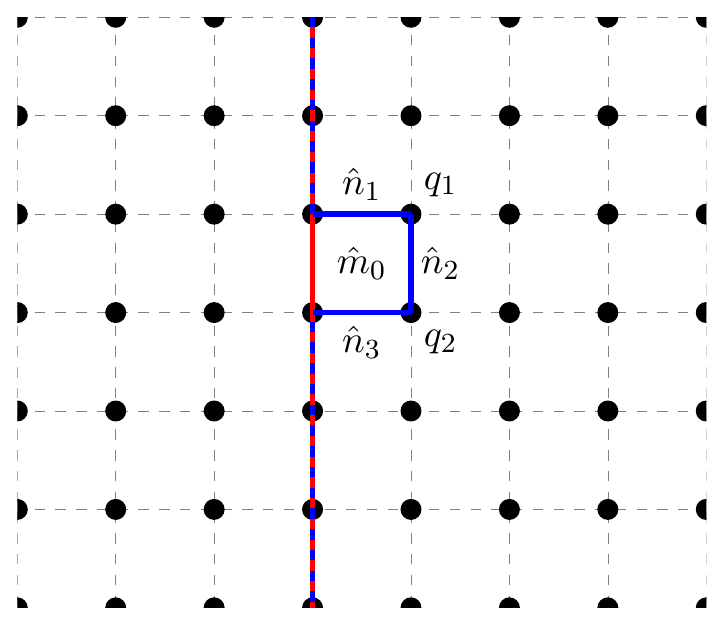}
    \caption{The red and blue interfaces are defined by the Dirichlet boundary conditions of the $\mathbb{Z}_N^{(0)}$ gauge fields. We can deform the red interface to the blue interface by integrating out $\hat m_0$ and gauge fixing $\hat n_1=\hat n_2=\hat n_3=0$. Keeping track of the appropriate normalization factors, we find that this deformation is topological.  Here we omit the superscripts that indicate the form degrees.}
    \label{fig:7}
\end{figure}

We can define an interface on the lattice by gauging the $\mathbb{Z}_N$ symmetry only in half of the spacetime.
We consider the picture as in Figure {\ref{fig:7}}.
To the left of the red line, we have the modified Villain XY-model (\ref{XY}).
To the right of the red line, we have the same theory coupled to a $\mathbb{Z}_N^{(0)}$ gauge theory (\ref{gaugedXY}).
Along the red line, we impose the Dirichlet boundary condition for the $\mathbb{Z}_N$ one-form gauge field, which amounts to setting 
\ie
\hat{n}^{(1)}| =q^{(0)}| = 0\,.
\fe

We now show that this interface defined by the Dirichlet boundary condition is topological, even with the coupling to the XY fields. 
That is, the value of the partition function doesn't change under small deformations of the interface. 
To show this, we consider a slightly deformed interface which we depicted as a blue line in Figure \ref{fig:7}.
We now compare the values of the two corresponding partition functions, $\mathcal Z_{\text{red}}$ and $\mathcal Z_{\text{blue}}$.

Compared to $\mathcal Z_{\text{blue}}$, $\mathcal Z_{\text{red}}$ includes more fields and more gauge parameters as labeled in Figure \ref{fig:7}. This leads to an additional normalization factor of 
\ie
\frac{1}{N^{\text{number of $\hat m$}}}\times \frac{1}{N^{\text{number of $q$}}}=\frac{1}{N}\times\frac{1}{N^2}~,
\fe
where this normalization factor is explained in Appendix \ref{app:topZN}. 

We can reduce $\mathcal Z_{\text{red}}$ to $\mathcal Z_{\text{blue}}$ by the following steps. 
Since the field $\hat m_0$ is not coupled to the matter fields, we can  integrate out the extra $\hat m_0$ in Figure \ref{fig:7}. This generates a factor of $N$ and constrains $\hat{n}_3 = -\hat{n}_1-\hat{n}_2$. 
Next, we set $\hat{n}_1 = \hat{n}_2 = 0$ using the gauge symmetry
\ie
\hat{n}_1 &\sim \hat{n}_1 + \hat{k}_1\,, \\
\hat{n}_2 &\sim \hat{n}_2 + \hat{k}_2 - \hat{k}_1\,.
\fe
There is a factor of $N^2$ coming from the volume of the gauge group that we gauge fixed. 
Putting everything together, the red interface is deformed into the blue one and the partition functions are related by
\begin{equation}
\mathcal Z_{\text{red}} = \frac{1}{N}\times\frac{1}{N^2} \times N \times N^2 \times \mathcal Z_{\text{blue}} = \mathcal Z_{\text{blue}}.
\end{equation}
We conclude that this interface in the modified Villain version of the XY model is topological. 
This essentially follows from the fact that the $\mathbb{Z}_N^{(0)}$ gauge field $\hat n^{(1)}$ is flat.

\subsubsection{Duality Defect as a Chern-Simons Coupling}\label{sec:XYlatticeD}

When $R=\sqrt{N}$, the theories on the two sides of the interface are isomorphic to each other. In this case, the topological interface defines a non-invertible topological defect, i.e. the duality defect as discussed in Section \ref{gaugingsec}, in a single theory. Consider the duality defect along the red line in Figure \ref{fig:7}. 
In order to describe the duality defect more explicitly, we perform the Poisson resummation \eqref{Possonresummationi} on the right-hand side of the interface. We start from the action (\ref{gaugedXY}) in the right half-space, gauge away $n$ and $\hat m$ using the gauge symmetries, and redefine the variables. In the right half-space,  this gives us
\begin{equation}
\frac{1}{4\pi N} \sum_{\text{link}} (\Delta \varphi^{(0)} - 2\pi \hat{n}^{(1)})^2
+i\sum_{\text{link}} \hat{n} ^{(1)}\Delta \tilde{\varphi}^{(0)} - i N\sum_{\text{defect}} n^{(1)} \tilde\varphi^{(0)},
\end{equation}
with the gauge symmetry \eqref{eq:gaugeafterdualize}. 
The third term is a boundary term localized along the defect.

\begin{figure}[ht]
    \centering
    \includegraphics[width=0.4\textwidth]{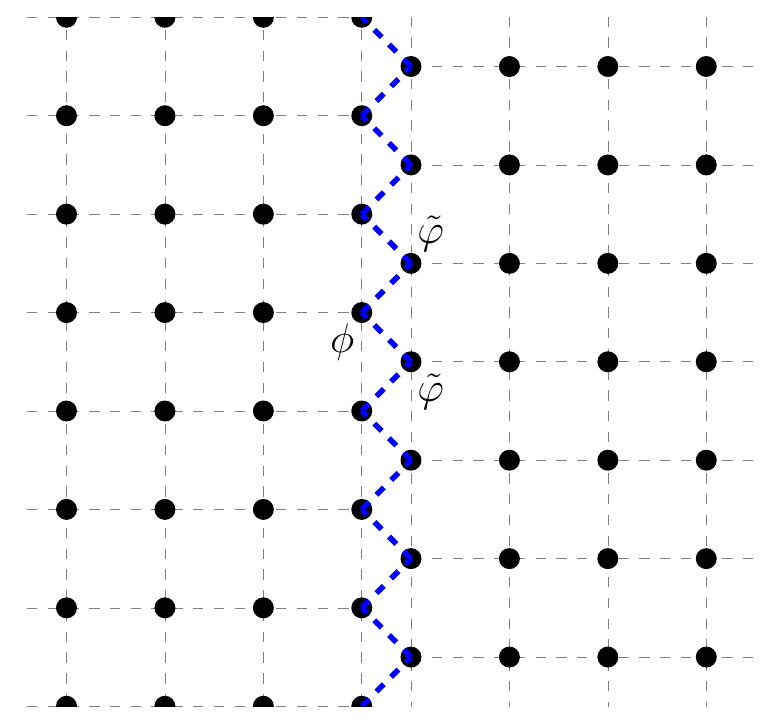}
    \caption{Duality defect is inserted along a cut between a lattice and its dual lattice. Fields along the cut are coupled by the blue dots as in  (\ref{defectcouple}).}
    \label{fig:8}
\end{figure}

Now, we perform the Poisson resummation \eqref{Possonresummationi} for the sum of $\hat{n}^{(1)}$ and obtain
\begin{equation}
\frac{N}{4\pi} \sum (\Delta \tilde{\varphi}^{(0)} - 2\pi \tilde{n}^{(1)})^2 - \frac{i}{2\pi} \sum \Delta \varphi^{(0)} (2\pi \tilde{n}^{(1)} - \Delta \tilde{\varphi}^{(0)}) - i N\sum_{\text{defect}} n^{(1)} \tilde\varphi^{(0)}. 
\end{equation}
Summing by parts in the second term then brings us to the theory at $R = \sqrt{N}$ defined on the dual lattice, plus a boundary term localized along the defect (as depicted in Figure \ref{fig:8}), which is given by
\begin{equation} \label{defectcouple}
-\frac{iN}{2\pi}\sum_{\text{defect}} \Big[\phi^{(0)}(\Delta \tilde{\varphi}^{(0)}-2\pi\tilde n^{(1)}) +  2\pi  n ^{(1)}\tilde\varphi^{(0)}\Big]\,,
\end{equation}
where we have used $\varphi^{(0)} = N\phi^{(0)}$ on the defect. 
In the continuum limit, it reduces to the 0+1d Chern-Simons coupling in the continuum action \eqref{contphi}.

\subsection{3+1d $U(1)$ Gauge Theory}

We next move on to the modified Villain lattice action of the 3+1d $U(1)$ gauge theory. 
Using the $S$-duality  that is realized explicitly on the lattice, we will construct the duality defect on the lattice.

\subsubsection{Modified Villain Model}

The theory is defined on a 4-dimensional Euclidean hypercube lattice.
The modified Villain action is given by
\begin{equation}
\frac{1}{g^2} \sum_{\text{2-cell}} (\Delta A^{(1)} - 2\pi n^{(2)})^2 + i \sum_{\text{3-cell}} \tilde{A}^{(1)} \Delta n^{(2)}.
\end{equation}
$A^{(1)}$ and $\tilde{A}^{(1)}$ are real-valued one-form gauge fields on the links and the dual links respectively, and $n^{(2)}$ is an integer-valued two-form gauge field on the plaquette.
We have the gauge transformations
\ie
A^{(1)} &\sim A^{(1)} + \Delta \alpha^{(0)} + 2\pi k^{(1)}\,, \nonumber \\
n^{(2)} &\sim n^{(2)} + \Delta k^{(1)}\,, \\
\tilde{A}^{(1)} &\sim \tilde{A}^{(1)} + \Delta \tilde{\alpha}^{(0)} + 2\pi \tilde{k}^{(1)}\,,
\fe
where $\alpha^{(0)}$ and $\tilde{\alpha}^{(0)}$ are real-valued gauge parameters on the sites and the dual sites, and $k^{(1)}$ and $\tilde{k}^{(1)}$ are integer-valued gauge parameters on the links and the dual links.
The gauge symmetry effectively makes $A^{(1)}$ and $\tilde{A}^{(1)}$ compact $U(1)$ gauge fields.

Analogous to the XY-model case, the theory exhibits an exact self-duality which can be seen by performing the Poisson resummation \eqref{Possonresummationi} for $n^{(2)}$ \cite{Sulejmanpasic:2019ytl,Gorantla:2021svj}.
Under the duality, ${4\pi}/{g^2} \leftrightarrow {g^2}/{4\pi}$, and $A^{(1)}$ and $\tilde{A}^{(1)}$ are exchanged.
This is the lattice version of the $S$-duality of the continuum $U(1)$ gauge theory.

The theory has a $U(1)_e^{(1)}$ electric one-form symmetry which shifts $A^{(1)}$ by a flat one-form gauge field.
One can gauge the $\mathbb{Z}_N^{(1)}$ subgroup of this symmetry.
The resulting theory is still a modified Villain lattice model of the $U(1)$ gauge theory, but with a new coupling $g'^2 =  g^2 N^2$.
Combined with the S-duality, we see that at the special point $4 \pi /g^2 = N$, the model is self-dual under this gauging.
Hence it admits the corresponding duality defect.

\subsubsection{Duality Defect as a Chern-Simons Coupling}\label{sec:U1latticeD}

We now focus on the theory at $4\pi/g^2=N$. The duality defect can be constructed by gauging the $\mathbb{Z}_N^{(1)}$ one-form symmetry in only half of the spacetime. Along the defect, we impose the topological boundary condition for the dynamical two-form field $\hat n^{(2)}\vert=0$. 

The Euclidean action for the whole system is
\ie
&\frac{N}{4\pi} \sum_{\text{2-cell}} (\Delta A_L^{(1)} - 2\pi n_L^{(2)})^2 + i \sum_{\text{3-cell}} \tilde{A}_L^{(1)} \Delta n_L^{(2)}
\\
&+
\frac{N}{4\pi} \sum_{\text{2-cell}} \left(\Delta \tilde B_R^{(1)} - 2\pi \tilde  n_R^{(2)}-\frac{2\pi}{N}\hat n_R^{(2)}\right)^2 + i \sum_{\text{3-cell}} {B}_R^{(1)} \Delta \left(\tilde n_R^{(2)}+\frac{1}{N}\hat n_R^{(2)}\right)+\frac{2\pi i}{N}\sum_{\text{3-cell}} \hat m_R^{(1)}\Delta\hat n_R^{(2)}.
\fe
Fields with subscript $L$ and $R$ live on the left-hand and the right-hand side of the duality defect, respectively. $A_L^{(1)},\tilde B_R^{(1)}$ and $\tilde A_L^{(1)},B_R^{(1)}$ are real-valued one-form gauge fields on links and dual links, respectively. $n_L^{(2)},\tilde n_R^{(2)},\hat n^{(2)}_R$ are integer-valued two-form gauge field on the plaquettes, and $\hat m_R^{(1)}$ is an integer-valued one-form gauge field on the dual links whose equation of motion constrains $\hat n_R^{(2)}$ to be a flat $\mathbb{Z}_N$ two-form gauge field. Along the duality defect, we have $A_L^{(1)}\vert=\tilde B_R^{(1)}\vert$, $n_L^{(2)}\vert=\tilde n_R^{(2)}\vert$ and $\hat n^{(2)}\vert=0$. 

The theory has the gauge symmetry
\begin{alignat}{2}
A_L^{(1)} &\sim A_L^{(1)} + \Delta \alpha_L^{(0)} + 2\pi k_L^{(1)}\,,\qquad\qquad && \tilde B_R^{(1)} \sim \tilde B_R^{(1)} + \Delta \tilde \alpha_R^{(0)} + 2\pi \tilde q_R^{(1)}+\frac{2\pi}{N} \tilde k_R^{(1)} \,, \nonumber
\\ 
n_L^{(2)} &\sim n_L^{(2)} + \Delta k_L^{(1)}\,,
&&\tilde n_R^{(2)} \sim \tilde n_R^{(2)} + \Delta \tilde q_R^{(1)}- \ell_R^{(2)}\,, \nonumber
\\
\tilde{A}_L^{(1)} &\sim \tilde{A}_L^{(1)} + \Delta \tilde{\alpha}_L^{(0)} + 2\pi \tilde{k}_L^{(1)}\,,
&&{B}_R^{(1)} \sim B_R^{(1)} + \Delta {\alpha}_R^{(0)} + 2\pi {q}_R^{(1)}\,,\nonumber
\\
\hat{n}_R^{(2)} &\sim \hat{n}_R^{(2)} + \Delta \tilde k_R^{(1)} + N\ell_R^{(2)}\,, &&
\hat m_R^{(1)} \sim \hat m_R^{(1)} - {q}_R^{(1)} + N k_R^{(1)}\,,
\end{alignat}
where $\alpha^{(0)}_L,\tilde\alpha_R^{(0)}$ and $\tilde\alpha_L^{(0)},\alpha_R^{(0)}$ are real-valued gauge parameters on the sites and the dual sites respectively, $k_L^{(1)},\tilde k_R^{(1)},\tilde q_R^{(1)}$ and $\tilde k_L^{(1)},k_R^{(1)},q_R^{(1)}$ are integer-valued gauge parameters on the links and dual links, and $\ell_R^{(1)}$ is an integer-valued gauge parameter on the plaquettes.

Using the gauge parameter ${q}_R^{(1)}$ and $\ell_R^{(2)}$, we can gauge fix $\hat m_R^{(1)}=0$ and $\tilde n_R^{(2)}=0$ except along the defects.  Next we redefine the compact $U(1)$ gauge fields $A_R^{(1)}= B_R^{(1)}/N$,  $\tilde A_R^{(1)}=N\tilde{B}_R^{(1)}$ and apply the Poisson resummation formula to the sum of $\hat n_R^{(2)}$. This gives us
\ie\label{eq:lattice_defect_U(1)}
&\frac{N}{4\pi} \sum_{\text{2-cell}} \left(\Delta A_L^{(1)} - 2\pi n_L^{(2)}\right)^2 + i \sum_{\text{3-cell}} \tilde{A}_L^{(1)} \Delta n_L^{(2)}
+\frac{N}{4\pi} \sum_{\text{2-cell}} \left(\Delta A_R^{(1)} - 2\pi n_R^{(2)}\right)^2 + i \sum_{\text{3-cell}} \tilde{A}_R^{(1)} \Delta n_R^{(2)}
\\
&+\frac{iN}{2\pi}\sum_{\text{defect}}\left[A_L^{(1)}\left(\Delta  A^{(1)}_R-2\pi n_R^{(2)}\right)-2\pi A_R^{(1)} n_L^{(2)}\right]
\fe
To the left and to the right of the duality defect, we have the modified Villain action of the $U(1)$ gauge theory on the lattice and the dual lattice, respectively. The fields on the two sides are coupled through the third term along the defect. In the continuum limit, this coupling  \eqref{eq:lattice_defect_U(1)} becomes the Chern-Simons coupling in \eqref{contU1}.

\subsection{3+1d $\mathbb{Z}_N$ Lattice Gauge Theory}\label{ZNsec}

In this subsection, we discuss the 3+1d $\mathbb{Z}_N$ lattice gauge theory in  the Villain formulation \cite{Elitzur:1979uv,Ukawa:1979yv,Gorantla:2021svj}. 
We review the Kramers-Wannier-like duality of the lattice model: the lattice model at weak coupling is dual to the one at strong coupling, but with an additional coupling to a topological $\mathbb{Z}_N^{(1)}$ two-form gauge theory. 
Therefore, at the self-dual coupling, the $\mathbb{Z}_N$ lattice model is invariant under gauging the $\mathbb{Z}_N^{(1)}$ one-form symmetry, which implies that there is a duality defect. 

In the case of the $\mathbb{Z}_2$ lattice gauge theory, we check that the expected phase at the self-dual point is consistent with our general theorems  in Section \ref{dynamicalsec}.

Our Villain lattice model can be viewed as a complementary lattice realization of the duality defect to \cite{Koide:2021zxj}.
More specifically,  the duality defect in our construction is realized as a Chern-Simons coupling between the $\mathbb{Z}_N$ gauge fields from the two sides.

\subsubsection{Villain Model}

We will follow closely the exposition in Appendix C.2 of \cite{Gorantla:2021svj}. 
The theory is defined on a 4-dimensional Euclidean hypercube lattice. On each link, there is an integer one-form gauge field $m^{(1)}$ and on each plaquette, there is an integer two-form gauge field $n^{(2)}$. The Villain action is
\ie\label{ZNpform}
\frac{ \beta}{2}\sum_{\text{plaquette}}\left(\Delta m^{(1)}-Nn^{(2)}\right)^2~.
\fe
It has the gauge symmetry
\ie
&m^{(1)}\sim m^{(1)}+\Delta \ell^{(0)}+Nk^{(1)}~,
\\
&n^{(2)}\sim n^{(2)}+\Delta k^{(1)}~,
\fe
where $\ell^{(0)}$ and $k^{(1)}$ are integer gauge parameters on the sites and the links respectively. The theory has an electric $\mathbb{Z}_N$ one-form global symmetry, which shifts $m^{(1)}$ by a flat integer one-form field.

In the limit $\beta\rightarrow\infty$, the theory becomes a topological $\mathbb{Z}_N^{(1)}$ lattice two-form gauge theory \cite{Dijkgraaf:1989pz,Kapustin:2014gua,Gorantla:2021svj}, described by the action
\ie\label{eq:ZNtopoaction}
\frac{2\pi i}{N}\sum_{\text{link}}m^{(1)}\Delta {\tilde n}^{(2)}~,
\fe
where ${\tilde n}^{(2)}$ is an integer-valued field with the integer gauge symmetry
\ie
&\tilde n^{(2)}\sim \tilde n^{(2)}+\Delta\tilde k^{(1)}+N\tilde q^{(2)}~.
\fe
Alternatively, it can also be viewed as an ordinary topological $\mathbb{Z}_N^{(0)}$ one-form gauge theory. 

We can dualize the theory \eqref{ZNpform} by performing Poisson resummation \eqref{Possonresummationi} to $n^{(2)}$. It leads to the action
\ie
\frac{4\pi^2}{2\beta N^2}\sum_{\text{plaquette}}\left(\tilde n^{(2)}\right)^2+\frac{2\pi i}{N}\sum_{\text{link}}m^{(1)}\Delta \tilde n^{(2)} ~,
\fe
where $\tilde n^{(2)}$ is an integer field on the dual plaquette.
We can introduce new gauge symmetries together with Stueckelberg fields, and write the action as
\ie
\frac{4\pi^2}{2\beta N^2}\sum_{\text{plaquette}}\left(\Delta\tilde m^{(1)}-N\hat n^{(2)}-\tilde n^{(2)}\right)^2+\frac{2\pi i}{N}\sum_{\text{link}}m^{(1)}\Delta \tilde n^{(2)} ~.
\fe
with the integer gauge symmetry
\ie
&\tilde m^{(1)}\sim \tilde m^{(1)}+\Delta\tilde\ell^{(0)}+\tilde k^{(1)}~,
\\
&\hat n^{(2)}\sim\hat n^{(2)}-\tilde q^{(2)}~,
\\
&\tilde n^{(2)}\sim \tilde n^{(2)}+\Delta\tilde k^{(1)}+N\tilde q^{(2)}~,
\\
&m^{(1)}\sim m^{(1)}+\Delta\ell^{(0)}+Nk^{(1)}~.
\fe
Thus, we see that the duality maps the 3+1d $\mathbb{Z}_N$ lattice gauge theory with coupling $\beta$ to another one  with $4\pi^2/\beta N^2$ that couples to  a topological $\mathbb{Z}_N^{(1)}$ two-form  gauge theory \eqref{eq:ZNtopoaction} \cite{Wegner:1971app,Savit:1979ny,Fradkin:1978th,Horn:1979fy,Elitzur:1979uv,Ukawa:1979yv}. (See also \cite{Kapustin:2014gua,Gorantla:2021svj} for recent discussions). 
The latter coupling is equivalent to gauging the $\mathbb{Z}_N^{(1)}$ one-form symmetry of the $\mathbb{Z}_N$ lattice gauge theory. 

In particular, at the self-dual point $\beta=2\pi/N$, the  lattice model is invariant under gauging the $\mathbb{Z}_N$ one-form global symmetry. 
From our discussion in Section \ref{gaugingsec}, it follows that the $\mathbb{Z}_N$ lattice gauge theory has a duality defect  at $\beta=2\pi/N$.

In Appendix \ref{app:hamiltonian}, we discuss the gauging of the one-form symmetry in the $\mathbb{Z}_2$ lattice gauge theory from a Hamiltonian formalism to arrive at a similar conclusion.

\subsubsection{Duality Defect as a Chern-Simons Coupling}

We now focus on the theory at the self-dual point $\beta=2\pi/N$. Following a similar analysis as in Sections \ref{sec:XYlatticeD} and \ref{sec:U1latticeD}, the duality defect, which divides the spacetime lattice into two halves, can be described as follows:
\ie
\frac{\pi}{N} \sum_{\text{plaquette}}\left(\Delta m_L^{(1)}-Nn_L^{(2)}\right)^2
+\frac{\pi}{N} \sum_{\text{plaquette}}\left(\Delta m_R^{(1)}-Nn_R^{(2)}\right)^2
+\frac{2\pi i}{N}\sum_{\text{defect}}m_L^{(1)}\Delta  m_R^{(1)}\,.
\fe
Here the subscripts $L,R$ indicate that the corresponding fields are on the left- or right-hand side of the duality defect. 
The lattice $\mathbb{Z}_N$ gauge theory is defined on the original lattice on the left-hand side of the duality defect,  and on the dual lattice on the other side. Along the defect, the fields from the two sides are coupled by a  Chern-Simons term.

To proceed, we first recall the theorem   in Section \ref{dynamicalsec}: 
Consider a general  system that is invariant under gauging a $\mathbb{Z}_2^{(1)}$ one-form symmetry. We proved that its low-energy phase (which we assumed is described by a relativistic QFT) cannot be trivially gapped. We now apply this theorem to the $\mathbb{Z}_2$ lattice gauge theory (which is a bosonic system). 

The $\mathbb{Z}_2$ lattice gauge theory has a phase transition between the confined and the deconfined phase. 
Numerics \cite{Creutz:1979zg} suggests that the transition is first-order, where the trivial, confined vacuum has the same energy as the vacuum supporting the continuum $\mathbb{Z}_2$ gauge theory.\footnote{We thank Meng Cheng for discussions on this point.}  
This is indeed consistent with our theorem as the IR phase is not trivially gapped.\footnote{Recently, it was numerically shown that the $\mathbb{Z}_2^{(1)}$ one-form symmetry at the 2+1d Ising model is spontaneously broken at the transition point \cite{Zhao:2020vdn}. This is consistent with the  expectation from the CFT at the second-order transition. Indeed,  we expect the disorder line to be described by a conformal line that exhibits perimeter law \cite{Billo:2013jda,Gaiotto:2013nva}.}

\section*{Acknowledgements}

We are grateful to M.\ Cheng, D. Freed, J.\ Kaidi, A.\ Kapustin, M.\ McLean, K.\ Ohmori, S.\ Seifnashri, C. Teleman, R.\ Thorngren, and Y.\ Wang for helpful conversations. 
We would also like to thank J.\ Kaidi, K. Ohmori, and Y. Zheng for the coordinated submission. CC is supported by the US Department of Energy DE-SC0021432 and the Simons Collaboration on Global Categorical Symmetries.  PSH is supported by the Simons Collaboration on Global Categorical Symmetries.  HTL is supported in part by a Croucher fellowship from the Croucher Foundation, the Packard Foundation and the Center for Theoretical Physics at MIT. 
The authors of this paper were ordered alphabetically. 

\appendix

\section{Topological Lattice $\mathbb{Z}_N^{(q)}$ Gauge Theory}\label{app:topZN}

Consider the $(q+1)$-form $\mathbb{Z}_N$ topological lattice gauge theory in the presentation of \cite{Gorantla:2021svj}:
\ie
S= \frac{2\pi i}{N}\sum_{(q+2)\text{-cell}} b^{(d-q-2)} \Delta a^{(q+1)} .
\fe
Here $a^{(q+1)}$ are $\mathbb{Z}_N$-valued gauge fields on the $(q+1)$-cells, and $b^{(d-q-2)}$ are $\mathbb{Z}_N$-valued gauge fields on the dual $(d-q-2)$-cells.

Define
\begin{equation}
\#i \equiv \text{number of $i$-cells} .
\end{equation}
The normalization for the partition function is 
\begin{equation}
{\cal Z} = \frac{1}{N^{\# (q+2)}} \times \frac{N^{(\# (q-1) +  \# (q-3) + \cdots)}}{N^{(\#q + \# (q-2) + \cdots)} } \sum_{\{a,b\}} \text{exp}\left(\frac{2\pi i}{N}\sum_{(q+2)\text{-cell}} b^{(d-q-2)} \Delta a^{(q+1)}\right).
\end{equation}
Changing the role of $a$ and $b$ in the normalization amounts to choosing a different Euler counterterm. 
On an arbitrary triangulated closed manifold $X^{(d)}$, with this normalization we reproduce the expected partition function
\begin{equation}
{\cal Z}[X^{(d)}] = \frac{|H^{q+1}(X^{(d)};\mathbb{Z}_N)| \times |H^{q-1}(X^{(d)};\mathbb{Z}_N)| \times \cdots}{|H^{q}(X^{(d)};\mathbb{Z}_N)| \times |H^{q-2}(X^{(d)};\mathbb{Z}_N)| \times \cdots}~.
\end{equation}
First, we integrate out $b^{(d-q-2)}$. It generates a factor of $N^{\# (q+2)}$ and constrains $a^{(q+1)}$ to be flat. Here we use the fact that the number of dual $(d-q-2)$-cells is the same as the number of $(q+2)$-cells.
Next, the sum over flat $a^{(q+1)}$ gives 
\ie
|H^{q+1}(X^{(d)};\mathbb{Z}_N)| \times (\text{\# of $(q+1)$-form pure gauge connections})\,.
\fe
The number of $(q+1)$-form pure gauge connections is equal to $\#q$ divided by the number of flat $q$-form connections that correspond to trivial gauge transformations. The flat $q$-form connections is then given by
\ie
|H^{q}(X^{(d)};\mathbb{Z}_N)| \times (\text{\# of $(q)$-form pure gauge connections})\,.
\fe
Iterating this argument we get the partition function.

\section{$\mathbb{Z}_2$ Lattice Gauge Theory in Hamiltonian Formalism}\label{app:hamiltonian}

In this appendix we discuss the 3+1d $\mathbb{Z}_2$ lattice gauge theory in the Hamiltonian formalism and gauge its $\mathbb{Z}_2$ one-form symmetry.  
It is complementary to the Lagrangian discussion in Section \ref{ZNsec}.

Consider the following Hamiltonian on a cubic lattice with a qubit on each link, acted by Pauli matrices $X_e,Y_e,Z_e$,
\begin{equation}
H=-U \sum_f \prod_{e\in f} Z_e -K\sum_e X_e-  \sum_v\prod_{e\ni v} X_e~,
\end{equation}
where the product in the first term is over all edges on the boundary of face $f$, and the product in the last term is over all edges connected to the vertex $v$.
The first and the second term can be respectively viewed as the squares of the magnetic and the electric fields, while the last term imposes the Gauss law energetically.

The theory has conserved charges supported on any closed surface $\tilde \Sigma$ on the dual lattice
\begin{equation}
Q(\tilde \Sigma)=\prod X_e~,
\end{equation}
where the product is over all edges cutting the surface.  
It generates a  one-form symmetry on the lattice.\footnote{To be more precise, the charge is not topological and depends on the shape of the surface $\tilde \Sigma$.  Only when the Gauss law is imposed strictly will the charge become a one-form symmetry operator \cite{Seiberg:2019vrp}. } 

In the limit $K=0$ the model is the ordinary toric code in 3+1d describing deconfined $\mathbb{Z}_2$ gauge theory. In the limit $K/U\rightarrow \infty$ the model becomes trivial Ising paramagnet. For finite coupling $K/U$ the model describes quantum phase transition(s) between confinement/deconfiment.

Let us gauge this one-form symmetry by introducing a new qubit on each face, acted by Pauli matrices $X_f,Y_f,Z_f$. The Gauss law is
\begin{equation}
X_e\prod_{f\ni e} X_f=1~,
\end{equation}
where the product is over all faces with boundary containing the edge $e$.
The first term in the original Hamiltonian is modified to $Z_f \prod Z_e$ to commute with the Gauss law constraint.
We can gauge-fix using the Gauss law constraint to obtain the
 Hamiltonian
\begin{equation}
H'=-U\sum_f Z_f -K\sum_e\prod_{f\ni e} X_f~,
\end{equation}
where the product in the last term is over all faces whose boundaries contain the edge $e$, and we have dropped 
the last term in the original Hamiltonian which is trivial using the Gauss law constraint. 
Moreover, we need to add flux term $-\sum_c \prod Z_f$ for the new gauge field to impose flat condition. Thus the new Hamiltonian after gauging is
\begin{equation}
\tilde H=-U\sum_f Z_f -K\sum_e\prod_{f\ni e} X_f-\sum_c \prod_{f\ni c} Z_f~,
\end{equation}
which is the same as the original Hamiltonian on the dual lattice with $U\leftrightarrow K$ and $X\leftrightarrow Z$.

The model is self-dual at $K=U$ under gauging the $\mathbb{Z}_2^{(1)}$ one-form symmetry, analogous to the Ising model in transverse field in 1+1d.
We conclude that the model at the self-dual point has a duality defect.

\bibliographystyle{JHEP}
\bibliography{ref}

\end{document}